%% file: acl_latex.tex
\documentclass[11pt]{article}

\usepackage[preprint]{acl}

\usepackage{times}
\usepackage{latexsym}
\usepackage{amssymb}
\usepackage[T1]{fontenc}

\usepackage[utf8]{inputenc}
\usepackage{tabularx}
\usepackage{microtype}

\usepackage{inconsolata}

\usepackage{graphicx}
\usepackage{booktabs}
\usepackage{multirow}
\usepackage{amsmath}
\usepackage{cleveref}
\usepackage{booktabs}
\usepackage{multirow}
\usepackage{tabularx}
\usepackage{ragged2e} 
\usepackage{makecell}

\title{
AQAScore: Evaluating Semantic Alignment in Text-to-Audio Generation via Audio Question Answering
}

\author{
Chun-Yi Kuan$^{\heartsuit}$, 
Kai-Wei Chang$^{\diamondsuit}$,
\and
Hung-yi Lee$^{\heartsuit}$ \\
$^{\heartsuit}$National Taiwan University \\
$^{\diamondsuit}$Massachusetts Institute of Technology
}

\begin{document}
\maketitle
\begin{abstract}

\input{sections/abstract}

\end{abstract}

\section{Introduction}

\input{sections/introduction}

\section{Related works}

\input{sections/related_works_v2}

\section{Methods}

\input{sections/methods}

\input{tables/benchmarks_intro}
\section{Experimental Setups}

\input{sections/experimental_setups}

\section{Experimental Results}

\input{tables/text_relevance}
\input{tables/pairwise_comparison_relate_baton}
\input{tables/composition_reasoning}

\input{sections/experimental_results}

\section{Discussion}

\input{sections/discussion}
\section{Conclusion}

\input{sections/conclusion}

\section*{Limitations}

\input{sections/limitation}

\section*{Ethics Statement}

\input{sections/ethics_statement}


\bibliography{anthology}

\appendix

\section{Appendix}
\label{sec:appendix}

\input{sections/appendix}

\end{document}

%% file: sections/abstract.tex

Although text-to-audio generation has made remarkable progress in realism and diversity, the development of evaluation metrics has not kept pace. 
Widely-adopted approaches, typically based on embedding similarity like CLAPScore, effectively measure general relevance but remain limited in fine-grained semantic alignment and compositional reasoning. 
To address this, we introduce AQAScore, a backbone-agnostic evaluation framework that leverages the reasoning capabilities of audio–aware large language models (ALLMs). 
AQAScore reformulates assessment as a probabilistic semantic verification task; rather than relying on open-ended text generation, it estimates alignment by computing the exact log-probability of a ``Yes'' answer to targeted semantic queries. 
We evaluate AQAScore across multiple benchmarks, including human-rated relevance, pairwise comparison, and compositional reasoning tasks. 
Experimental results show that AQAScore consistently achieves higher correlation with human judgments than similarity-based metrics and generative prompting baselines, showing its effectiveness in capturing subtle semantic inconsistencies and scaling with the capability of underlying ALLMs.

%% file: sections/introduction.tex
Recent advances in text-to-audio (TTA) generation~\cite{liu2023audioldm, audioldm2-2024taslp, majumder2024tango, ghosal2023tango, hung2024tangofluxsuperfastfaithful, leeetta,
huangimpact, evans2025stable, yang2023diffsound} have enabled models to produce increasingly realistic, diverse, and semantically rich audio from natural language descriptions.
These models have opened up new possibilities in sound design, accessibility, and multimedia creation, marking rapid progress similar to that seen earlier in text-to-image generation.
However, evaluation methods for text relevance, defined as the alignment between the generated audio and the input prompt, have not kept pace with the rapid advances in generative modeling. 
Widely-adopted metrics, such as CLAPScore~\cite{wu2023large, elizalde2023clap}, rely on embedding similarity between audio and text.
While effective for measuring coarse relevance, these metrics are limited in assessing whether the generated audio truly conveys the meaning expressed in the text, particularly when the description involves multiple sound events, temporal order, or fine-grained attributes.
Prior analyses~\cite{ghoshcompa} have shown that CLAP-based similarity performs poorly on compositional reasoning aspects, struggling to distinguish semantically inverted or attribute-swapped audio–text pairs (e.g., ``a dog barking before thunder'' vs. ``thunder before a dog barking'').
These findings suggest that improving text–audio evaluation requires a deeper understanding of how meaning is structured and composed within sounds.
To bridge this gap, we explore evaluation methods that move beyond simple embedding similarity. We leverage the audio question answering capabilities of modern audio-aware large language models (ALLMs), which support more fine-grained and controllable assessments of audio-text alignment.

\input{figures/overview}

To this end, we propose \textbf{AQAScore}, an evaluation framework that reformulates text-to-audio assessment as an audio-to-text understanding task.
We illustrate the framework in Figure~\ref{fig:overview}.
Leveraging the recent progress of ALLMs~\cite{ghoshaudio, xu2025qwen2, arora2025on}, which exhibit strong auditory comprehension and reasoning abilities, we employ them as the backbone of an audio question answering (AQA) model.
Given an audio sample and a text description, the model is asked a question such as \texttt{``Does this audio contain the sound events described by the text: \{description\}?''}; the probability of a \texttt{``Yes''} response provides a continuous measure of semantic consistency between the two modalities.
In contrast to similarity-based metrics that rely on feature proximity, AQAScore assesses alignment by calculating the probability that the audio contains the specific sound events described in the text. 
This allows it to detect fine-grained semantic inconsistencies, including missing or incorrect elements that global similarity scores often overlook~\cite{alhamoud2025vision, kamath2024hard, yuksekgonuland, awal2024vismin}.

We evaluate AQAScore across three complementary settings to systematically assess both its perceptual alignment and reasoning capability:
(1) Human-rated relevance benchmarks such as RELATE~\cite{kanamori2025relate} and PAM~\cite{deshmukh2024pam} to measure consistency with human judgments,
(2) Pairwise comparison benchmarks, including Baton~\cite{liao2024baton} and a pairwise version of RELATE~\cite{kanamori2025relate}, where the original RELATE dataset is converted into a pairwise setting to evaluate the model’s ability to reproduce human preferences, and
(3) Compositional reasoning benchmarks like CompA~\cite{ghoshcompa} to analyze its understanding of fine-grained semantics including event order and attributes.
Our results show that AQAScore correlates more strongly with human judgments and can also capture semantic inconsistencies that similarity-based metrics often overlook.

In summary, our contributions are as follows:
(1) We introduce AQAScore, a evaluation framework that reformulates text–audio alignment as an audio question answering task.
(2) We evaluate AQAScore through human-rated relevance, pairwise comparison, and compositional reasoning benchmarks, showing higher correlation with human judgments and greater sensitivity to semantic errors.

%% file: figures/overview.tex
\begin{figure}[t]
    \centering
    \includegraphics[width=0.48\textwidth]{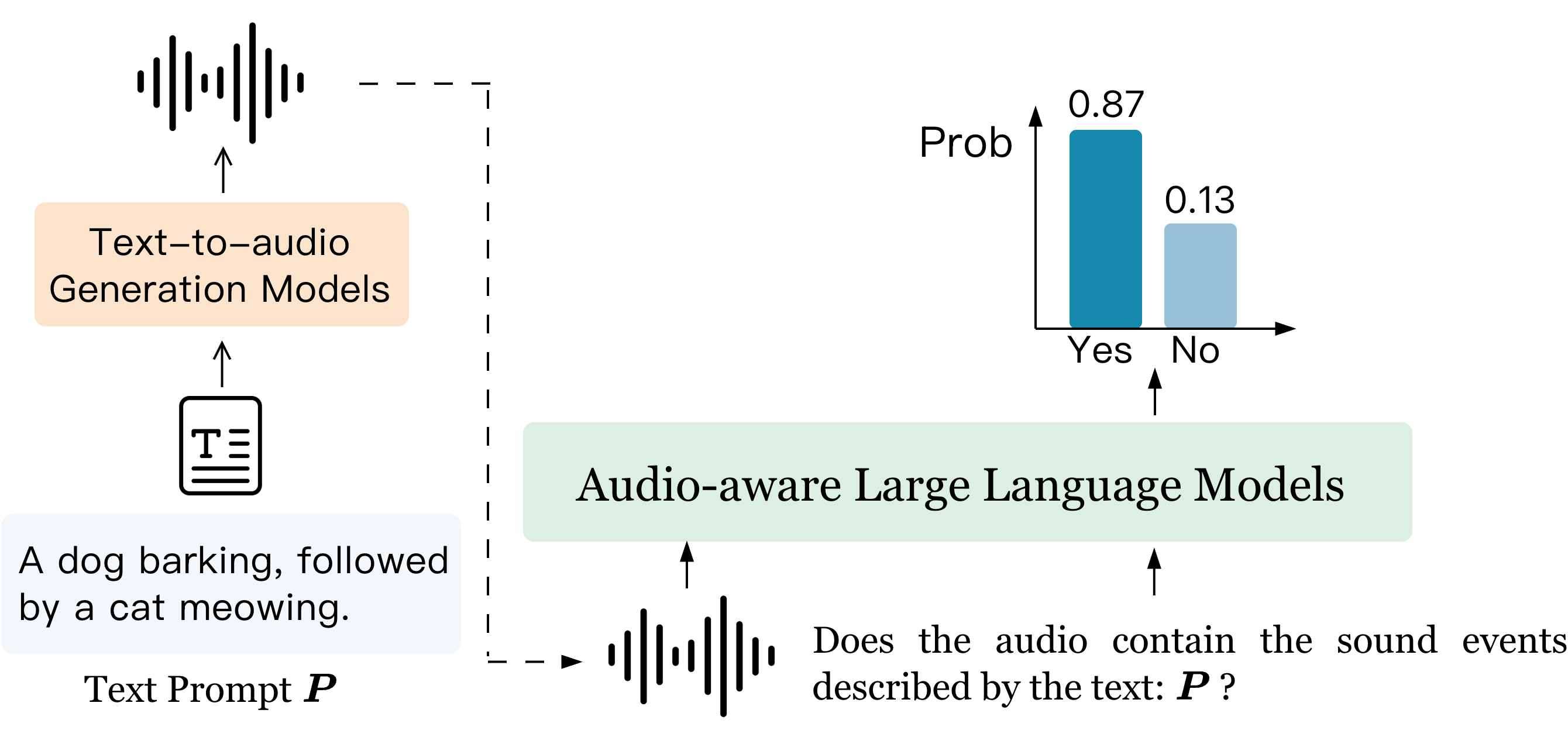}
    \caption{Overview of AQAScore framework.}
    \label{fig:overview}
\end{figure}

%% file: sections/related_works_v2.tex
Recent advancements in text-to-audio (TTA) generation~\cite{audioldm2-2024taslp, majumder2024tango, ghosal2023tango, kuan2023towards, hung2024tangofluxsuperfastfaithful, leeetta, huangimpact, evans2025stable} have achieved impressive progress in producing realistic and diverse audio content. 
Models such as Tango2~\cite{majumder2024tango} leverage latent diffusion or large-scale audio–language pretraining to synthesize complex soundscapes aligned with textual descriptions. 
Despite advances in generation quality, assessing whether the produced audio faithfully reflects the intended semantics remains challenging.
Existing evaluation methods largely rely on embedding similarity metrics such as CLAPScore, which capture global relevance but struggle with structured reasoning, event ordering, and attribute binding. 
While human-annotated benchmarks like Baton~\cite{liao2024baton}, PAM~\cite{deshmukh2024pam}, and RELATE~\cite{kanamori2025relate} provide reliable human judgments to address these complexities, their scalability is limited by high costs, motivating a shift toward model-based evaluation.

In the visual domain, VQAScore~\cite{lin2024evaluating} reformulates image–text evaluation as a question answering problem, showing that reasoning-capable multimodal LLMs can act as effective evaluators beyond simple score prediction.
In the audio and speech domains, audio-aware large language models~\citep{xu2025qwen2, ghoshaudio, kuan2025teaching, kuan2025alignment, abouelenin2025phi, fathullah2023towards} (ALLMs) have increasingly been explored as judges for various evaluation tasks, including speech quality assessment~\cite{chenaudio}, paralinguistic analysis~\cite{manakul2025audiojudge}, speaking style evaluation~\cite{chiang2025audiojudge}, instruction-following in text-to-speech systems~\cite{huang2025instructttseval}, and dialogue temporal dynamics~\cite{chang2025game}. 
For TTA generation specifically, recent works such as AudioEval~\cite{wang2025audioeval} and TTA-Bench~\cite{wang2025tta} have demonstrated the potential of learned evaluators for system-level and perceptual comparisons. 
Driven by large-scale audio–language alignment training~\cite{xu2025qwen2, ghoshaudio}, recent ALLMs have evolved beyond high-level perceptual assessment and can now directly answer complex textual queries about audio content. 
While this capability enables more fine-grained evaluation, existing protocols predominantly treat ALLMs as generative judges, relying on free-form descriptions or discrete ratings. 
However, we observe that such generative prompting approaches struggle to achieve strong correlation with human judgments when assessing audio–text consistency.

To address this, we introduce AQAScore, which reformulates evaluation as a probabilistic verification task. Rather than relying on open-ended text generation, AQAScore extracts the exact log-probability of a ``Yes'' response to targeted semantic queries. Our experiments demonstrate that this probabilistic approach provides a more sensitive and continuous metric that aligns more closely with human perception than direct generative prompting, offering a more robust framework for evaluating semantic alignment in text-to-audio generation.

%% file: sections/methods.tex
Given an audio–text pair $(a, t)$, we construct a question 
$q(t)$ such as \texttt{``Does this audio contain the sound events described by the text: \{t\}? Please answer yes or no.''}
Here, $t$ denotes the text description that serves as the input prompt for text-to-audio generation.
Let $p_\theta(y \mid a, q(t))$ denote the conditional probability 
predicted by an audio question answering (AQA) model 
parameterized by $\theta$, where $y \in \{\text{``Yes''}, \text{``No''}\}$. 
We compute the log-likelihoods of the two responses as:
\[
\begin{aligned}
s_{\text{yes}} &= \log p_\theta(\text{``Yes''} \mid a, q(t)), \\
s_{\text{no}}  &= \log p_\theta(\text{``No''} \mid a, q(t)).
\end{aligned}
\]
We define the \textbf{AQAScore} as the softmax-normalized probability of the ``Yes'' token relative to the ``No'' token:
\[
\operatorname{AQAScore}(a, t) = \frac{\exp(s_{\text{yes}})}{\exp(s_{\text{yes}}) + \exp(s_{\text{no}})}
\]
The resulting score can be interpreted as the model’s confidence that the audio semantically satisfies the textual description, providing a continuous measure of alignment between the two modalities.

For pairwise comparison scenarios, AQAScore can be extended to assess relative preferences between audio or text candidates.
Given two audio samples $(a_1, a_2)$ and a common textual description $t$, 
the model predicts which audio better aligns with the text by comparing their individual scores:
\[
\mathrm{Pref}(a_1, a_2, t)
= \arg\max_{a_i \in \{a_1, a_2\}}
   \operatorname{AQAScore}(a_i, t)
\]

This setup corresponds to evaluating text-to-audio generation systems under a shared text prompt.
Conversely, when a single audio $a$ is paired with two textual descriptions $(t_1, t_2)$, 
the preference is determined by comparing $\operatorname{AQAScore}(a, t_1)$ and $\operatorname{AQAScore}(a, t_2)$, 
reflecting which caption better describes the same audio:
\[
\mathrm{Pref}(a, t_1, t_2)
= \arg\max_{t_i \in \{t_1, t_2\}}
   \operatorname{AQAScore}(a, t_i)
\]
In both cases, the higher AQAScore indicates stronger semantic consistency, 
and the predicted preference is later compared with human annotations to compute accuracy.



%% file: tables/benchmarks_intro.tex
\begin{table}[!htbp]
\centering
\small
\caption{Summary of benchmarks and evaluation tasks.}
\label{tab:benchmarks-intro}

\newcolumntype{L}[1]{>{\RaggedRight\hsize=#1\hsize\arraybackslash}X}
\setlength{\tabcolsep}{6pt}
\renewcommand{\arraystretch}{1.3}

\begin{tabularx}{\columnwidth}{@{} L{0.8} L{1.0} L{1.2} @{} }
\toprule
\textbf{Category} & \textbf{Dataset} & \textbf{Evaluation Task} \\ \midrule

\multirow{5}{=}{Audio-Text Relevance} 
    & RELATE, PAM & Alignment with human subjective ratings. \\ 
    \cmidrule(l{3pt}r{3pt}){2-3} 
    & Baton & Alignment with binary human preferences. \\ \midrule

Pairwise Comparison & RELATE-Pair, Baton-Pair & Pairwise preference judgment of audio clips. \\ \midrule

Compositional Reasoning & CompA-Order, CompA-Attribute & Temporal and attribute reasoning. \\ \bottomrule

\end{tabularx}
\end{table}

%% file: sections/experimental_setups.tex
\subsection{Backbone Models for AQAScore}
In this study, we use the Qwen2.5-Omni~\cite{xu2025qwen2} models (3B and 7B) and the 7B Audio Flamingo 3~\cite{ghoshaudio} model. 
For Audio Flamingo 3, we refer to the base, chat, and think versions as AF3, AF3-Chat, and AF3-Think, respectively. 
We select these models because they are open-source, widely adopted, and demonstrate strong performance on audio understanding and reasoning benchmarks~\cite{sakshimmau, kuan2024understanding, kuan2025can, huang2024dynamic, huangdynamic2025, kuan2026}.
\subsection{Benchmark Datasets}
We evaluate AQAScore across three categories of benchmarks, ranging from human relevance ratings to compositional reasoning tasks. 
A summary of these datasets is provided in Table~\ref{tab:benchmarks-intro}.

\noindent \textbf{Human-rated Text Relevance.}
We adopt the RELATE~\citep{kanamori2025relate} and PAM~\citep{deshmukh2024pam} benchmarks to evaluate text–audio semantic alignment against human judgments. 
RELATE is a subjective evaluation dataset designed for assessing the relevance between textual descriptions and corresponding audio clips generated by recent text-to-audio models such as AudioLDM2 and Tango2. 
It provides human-rated scores ranging from 0 to 10 as ground truth for semantic alignment, with scores binned into 2-point intervals.
We use the test split of RELATE for evaluation and average the human ratings for each instance. 
On the other hand, PAM includes human preference ratings (from 1 to 5) on text–audio relevance, serving as another benchmark for correlation analysis with human perception.

\noindent \textbf{Pairwise Comparison.}
We evaluate model performance using a pairwise comparison protocol derived from the RELATE~\cite{kanamori2025relate} and Baton~\cite{liao2024baton} benchmarks.
First, we convert the RELATE dataset into a pairwise format.
For each text prompt, we rank audio clips generated by different models according to human ratings and construct audio pairs to test whether the model can correctly identify which clip better matches the text description.
To ensure clear preference signals, we only include pairs where the human scores differ by more than two points.
We refer to this converted dataset as \textit{RELATE-Pair}.

Second, we use the Baton~\cite{liao2024baton} dataset, which contains human-curated text prompts and corresponding generated audio annotated with binary human feedback (preferred or rejected).
This allows us to directly evaluate whether model preferences align with human judgments.
In addition, Baton provides two types of text instructions: one describing two sound events and the other describing three sound events, corresponding to different levels of difficulty.
We further convert Baton into a pairwise comparison setting by pairing, for the same text prompt, one human-preferred audio clip with one human-rejected clip.
The resulting pairs are used to test whether the model can correctly identify the audio that better matches the text description.
We refer to this converted dataset as \textit{Baton-Pair}.

\noindent \textbf{Compositional Reasoning.}  
We employ the CompA benchmark~\cite{ghoshcompa} to evaluate our model’s ability in audio compositional reasoning.  
CompA is a diagnostic benchmark designed to test how well audio–language models (ALMs) can reason about the composition of acoustic events.  
It comprises two expert-annotated sub-benchmarks:  
CompA-order evaluates the model’s understanding of the \emph{order or occurrence} of acoustic events in an audio clip.
For example, models must distinguish between captions such as ``A dog barks before a car engine starts'' and ``A car engine starts before a dog barks.''
CompA-attribute evaluates the model’s ability to bind \emph{attributes} to specific acoustic events.
For instance, models need to tell apart ``A woman laughs while a baby cries'' and ``A baby laughs while a woman cries.''
The benchmark reports three complementary metrics: \textit{Text Score}, \textit{Audio Score}, and \textit{Group Score}.
The Text Score measures whether a model can select the correct caption when given an audio clip, and the Audio Score tests whether the model can identify the correct audio when given a caption.
Finally, the Group Score combines both criteria, requiring the model to succeed in both directions simultaneously.
Together, these metrics offer a more complete assessment of an ALM’s compositional reasoning ability, balancing text–audio retrieval consistency and overall robustness.

\subsection{Baseline Methods}
To ensure a fair comparison, we include several baseline evaluation approaches. 
\textbf{CLAPScore} serves as a widely adopted similarity-based metric that measures the embedding similarity between audio and text representations:
\[
\operatorname{CLAPScore}(a, t)
= \frac{f_\text{a}(a) \cdot f_\text{t}(t)}
       {\|f_\text{a}(a)\| \, \|f_\text{t}(t)\|},
\]
where $f_\text{a}$ and $f_\text{t}$ denote the CLAP audio and text encoders, respectively.
Here, $a$ and $t$ refer to an audio–text pair used for evaluating cross-modal alignment.
Higher scores indicate stronger embedding-level alignment between the modalities.
CLAPScore results are reported using several pretrained checkpoints (see Appendix~\ref{appendix:clap-checkpoints} for full checkpoint names and details).

In addition, we include a \textbf{direct prompting} baseline, where ALLMs are directly prompted to either rate the text–audio relevance on a Likert scale or select the more relevant audio (or text) in pairwise comparison scenarios.
Note that for the rating setup, the instruction-guided ALLM follows the dataset-specific annotation guidelines provided in the corresponding original papers.
Building on this, and inspired by the cascade pipeline in~\citet{kuan2024speech}, we further include a \textbf{cascade} two-stage baseline.
In the first stage, the ALLM is prompted to produce a detailed description of the audio, and in the second stage, the model is asked whether this description matches the target text and rate the text–audio relevance on a Likert scale.
The details and full prompt templates used in our experiments are listed in Appendix~\ref{appendix:full-prompt-templates}.

\subsection{Evaluation Metrics}

To measure the effectiveness of each automatic metric, we compute its correlation with human judgments across benchmark datasets. 
Specifically, we report three standard correlation measures: Pearson (PCC), Spearman (SRCC), and Kendall’s $\tau$.
PCC evaluates the linear relationship between the automatic scores and human ratings, 
while SRCC and Kendall's $\tau$ assess monotonic rank consistency, which is particularly suitable when the absolute rating scale may vary across evaluators. 
For the prompting and cascade baselines, we correlate the model-predicted scores (e.g., Likert-scale ratings) with human ratings using the same coefficients.

For datasets in a pairwise comparison format like Baton-Pair and RELATE-Pair, 
we measure the percentage of correctly predicted preferences, 
i.e., the proportion of pairs where the metric assigns a higher score to the same sample preferred by humans.
For the prompting and cascade baselines under pairwise settings, 
we instead use the model’s predicted preference labels (e.g., which of the two audios is more relevant) 
and compute accuracy by comparing them against human preference annotations.
In addition, since the original Baton dataset provides text–audio pairs with binary human feedback (preferred or rejected), we compute the receiver operating characteristic (ROC) curve and the corresponding area under the curve (AUC) to measure how well the model’s scores align with human preferences.

%% file: tables/text_relevance.tex
\begin{table*}[ht]
\centering
\small
\begin{tabular}{llccccccc}
\toprule
\textbf{Methods} & \textbf{Models} & \multicolumn{3}{c}{RELATE~\citep{kanamori2025relate}} & \multicolumn{3}{c}{PAM~\citep{deshmukh2024pam}} \\
\cmidrule(lr){3-5} \cmidrule(lr){6-8}
 &  & LCC $\uparrow$ & SRCC $\uparrow$ & KTAU $\uparrow$
 & LCC $\uparrow$ & SRCC $\uparrow$ & KTAU $\uparrow$ \\
\midrule
\multicolumn{8}{l}{\textit{Baseline Methods~\citep{kanamori2025relate}}} \\
RELATE\textsuperscript{\dag}
& – 
& 0.385 & 0.383 & 0.265
& – & – & – \\
RELATE w/ CBL\textsuperscript{\dag} 
& – 
& 0.377 & 0.374 & 0.259
& – & – & – \\
\midrule
\multirow{5}{*}{AQAScore} 
& Qwen2.5-Omni-3B 
& 0.443 & 0.453 & 0.327
& \underline{0.540} & 0.560 & \underline{0.410} \\
& Qwen2.5-Omni-7B 
& \textbf{0.544} & \textbf{0.556} & \textbf{0.396}
& 0.518 & \textbf{0.589} & \textbf{0.429} \\
& AF3 
& \underline{0.475} & \underline{0.508} & \underline{0.357}
& 0.496 & 0.538 & 0.383 \\
& AF3-Think 
& 0.435 & 0.474 & 0.330
& \textbf{0.582} & \underline{0.587} & 0.419 \\
& AF3-Chat 
& 0.300 & 0.389 & 0.271
& 0.381 & 0.435 & 0.337 \\
\midrule
\multirow{5}{*}{CLAPScore} 
& Music+AS 
& 0.379 & 0.364 & 0.248 
& 0.436 & 0.436 & 0.303 \\
& 630k-AS 
& 0.442 & 0.420 & 0.289
& 0.384 & 0.377 & 0.261 \\
& Music+Speech+AS 
& 0.369 & 0.355 & 0.244
& 0.435 & 0.434 & 0.301 \\
& Music+Speech 
& 0.395 & 0.351 & 0.241
& 0.463 & 0.454 & 0.317 \\
& 630k-best 
& 0.448 & 0.432 & 0.291
& 0.472 & 0.477 & 0.337 \\
\midrule
\multirow{8}{*}{Prompting} 
& Qwen2.5-Omni-3B
& 0.290 & 0.260 & 0.215
& 0.408 & 0.367 & 0.296 \\
& Qwen2.5-Omni-7B
& 0.435 & 0.397 & 0.322
& 0.450 & 0.395 & 0.317 \\
& AF3
& 0.368 & 0.361 & 0.296
& 0.235 & 0.210 & 0.172 \\
& AF3-Think
& 0.144 & 0.122 & 0.100
& 0.053 & 0.053 & 0.044 \\
& AF3-Chat
& 0.152 & 0.155 & 0.125
& 0.041 & 0.037 & 0.030 \\
& GPT-4-Audio 
& 0.300 & 0.347 & 0.279
& 0.209 & 0.186 & 0.152 \\
& Gemini-2.5-Flash 
& 0.331 & 0.302 & 0.242
& 0.286 & 0.231 & 0.187 \\
& Gemini-2.5-Pro 
& 0.397 & 0.388 & 0.309
& 0.330 & 0.293 & 0.232 \\
\midrule
\multirow{5}{*}{Cascade} 
& Qwen2.5-Omni-3B
& 0.369 & 0.373 & 0.288
& 0.425 & 0.428 & 0.333
\\
& Qwen2.5-Omni-7B
& 0.405 & 0.409 & 0.319
& 0.432 & 0.429	& 0.336
\\
& AF3
& 0.285 & 0.296 & 0.228
& 0.474 & 0.465 & 0.361
\\
& AF3-Think
& 0.168	& 0.169 & 0.132
& 0.024 & 0.030 & 0.025
\\
& AF3-Chat
& 0.240 & 0.236 & 0.184
& 0.410 & 0.410 & 0.319
\\
\bottomrule
\end{tabular}
\caption{
Performance on the RELATE and PAM benchmarks.
``–'' indicates results not available.
$\dagger$ denotes that we directly use the best baseline results reported in the original papers.
Bold indicates the best performance, and underline indicates the second best.
}
\label{tab:relate}
\end{table*}


%% file: tables/pairwise_comparison_relate_baton.tex
\begin{table*}[ht]
\centering
\small
\begin{tabular}{llccccc}
\toprule
\textbf{Methods} & \textbf{Models} 
& \multicolumn{1}{c}{\textbf{RELATE-Pair}}
& \multicolumn{2}{c}{\textbf{Baton}} 
& \multicolumn{2}{c}{\textbf{Baton-Pair}}
\\
\cmidrule(lr){3-3} \cmidrule(lr){4-5} \cmidrule(lr){6-7}
 &  
 & Pair Acc. $\uparrow$
 & AUC (2) $\uparrow$
 & AUC (3) $\uparrow$
 & Pair Acc. (2) $\uparrow$ 
 & Pair Acc. (3) $\uparrow$
 \\
\midrule

\multirow{5}{*}{AQAScore} 
& Qwen2.5-Omni-3B 
& 75.0
& \underline{0.68} & 0.61 & \underline{70.3} & \textbf{69.1} \\
& Qwen2.5-Omni-7B
& \textbf{77.6}
& \textbf{0.69} & \textbf{0.65} & \textbf{71.0} & \underline{67.8} \\
& AF3
& 73.6
& 0.64 & 0.60 & 66.3 & 63.2 \\
& AF3-Think
& 72.4
& 0.66 & 0.61 & 65.8 & 62.0 \\
& AF3-Chat
& 72.1
& 0.57 & 0.54 & 63.1 & 60.2 \\
\midrule

\multirow{5}{*}{CLAPScore} 
& Music+AS
& \underline{76.0}
& 0.57 & 0.56 & 65.8 & 60.0 \\
& 630k-AS
& 75.2
& 0.53 & 0.62 & 64.6 & 57.5 \\
& Music+Speech+AS
& 69.8
& 0.56 & 0.54 & 64.7 & 58.0 \\
& Music+Speech
& 71.0
& 0.55 & \underline{0.63} & 66.2 & 56.8 \\
& 630k-best
& 71.0
& 0.58 & 0.62 & 69.0 & 61.9 \\
\midrule

\multirow{5}{*}{Prompting} 
& Qwen2.5-Omni-3B
& 66.2
& 0.50 & 0.50 & 50.3 & 50.1 \\
& Qwen2.5-Omni-7B
& 61.0
& 0.50 & 0.50 & 55.8 & 54.3 \\
& AF3
& 44.5
& 0.50 & 0.50 & 50.6 & 50.7 \\
& AF3-Think
& 41.7
& 0.50 & 0.50 & 49.8 & 50.0 \\
& AF3-Chat
& 63.3
& 0.50 & 0.50 & 54.6 & 50.5 \\
\midrule

\multirow{5}{*}{Cascade} 
& Qwen2.5-Omni-3B
& 68.6
& 0.52 & 0.55 & 37.3 & 35.6 \\
& Qwen2.5-Omni-7B
& 67.1
& 0.59 & 0.53 & 46.1 & 38.2 \\
& AF3
& 64.5
& 0.50 & 0.50 & 43.3 & 38.0 \\
& AF3-Think
& 55.6
& 0.50 & 0.50 & 15.6 & 22.6 \\
& AF3-Chat
& 60.0
& 0.50 & 0.51 & 45.5 & 42.0 \\
\bottomrule
\end{tabular}

\caption{
Performance comparison on the RELATE-Pair, Baton, and Baton-Pair benchmarks.
For Baton and Baton-Pair, (2) and (3) denote evaluation under two-sound-event and three-sound-event label settings, respectively.
}
\label{tab:relate-baton-pairwise}
\end{table*}

%% file: tables/composition_reasoning.tex
\begin{table*}[ht]
\centering
\small
\begin{tabular}{llcccccc}
\toprule
\textbf{Methods} & \textbf{Models} & \multicolumn{3}{c}{CompA-Order} & \multicolumn{3}{c}{CompA-Attribute} \\
\cmidrule(lr){3-5} \cmidrule(lr){6-8}
 &  & Text $\uparrow$ & Audio $\uparrow$ & Group $\uparrow$ 
 & Text $\uparrow$ & Audio $\uparrow$ & Group $\uparrow$
 \\
\midrule
\multirow{5}{*}{AQAScore} 
& Qwen2.5-Omni-3B 
& 50.00 & 35.00 & 25.25 
& 29.95 & 23.35 & 14.21 \\
& Qwen2.5-Omni-7B 
& \textbf{67.00} & \textbf{52.00} & \textbf{45.75} 
& \underline{44.16} & \textbf{37.06} & \textbf{24.87} \\
& AF3 
& \underline{51.00} & \underline{42.50} & 32.25
& 41.33 & \underline{33.67} & \underline{18.88} \\
& AF3-Think 
& 42.25 & 27.75 & 17.00 
& 33.67 & 22.45 & 11.73 \\
& AF3-Chat 
& 24.00 & 28.00 & 14.00 
& 17.86 & 14.80 & 4.08 \\
\midrule
\multirow{6}{*}{CLAPScore} 
& CompA-CLAP 
& 40.70 & 35.60 & \underline{33.85} 
& \textbf{44.28} & 22.52 & 15.13 \\
& Music+AS 
& 14.75 & 7.50 & 2.25 
& 37.06 & 11.17 & 9.14 \\
& 630k-AS 
& 19.50 & 6.50 & 3.00 
& 34.52 & 5.58 & 4.06 \\
& Music+Speech+AS 
& 15.25 & 4.50 & 1.75 
& 30.96 & 11.17 & 8.63 \\
& Music+Speech 
& 17.50 & 5.25 & 2.50 
& 34.52 & 9.14 & 7.61 \\
& 630k-best 
& 21.50 & 9.25 & 4.50 
& 35.53 & 17.26 & 11.17 \\
\bottomrule
\end{tabular}
\caption{Results on the CompA benchmark for the \textit{Order} and \textit{Attribute} sub-tasks.
Bold and underline denote the best and second-best results, respectively (\%).
}
\label{tab:compa}
\end{table*}


%% file: sections/experimental_results.tex
\subsection{Human-rated Text Relevance}

We begin by examining whether AQAScore aligns with human perception of text–audio relevance. 
As presented in Table~\ref{tab:relate}, for both the RELATE and PAM datasets, AQAScore based on Qwen2.5-Omni-7B generally achieves higher correlation with human ratings (in terms of Pearson, Spearman, and Kendall coefficients) than CLAPScore and other baseline methods. 
This indicates that using ALLMs as audio question answering evaluators better captures perceptual alignment between audio and text. 
For fair comparison, we also include the baselines reported in the original RELATE papers~\cite{kanamori2025relate}. 
These baselines are trained on the RELATE training set, whereas both AQAScore and CLAPScore operate in a zero-shot manner without any dataset-specific tuning.

Because we cannot access the log probabilities of proprietary models such as GPT-4-Audio~\cite{hurst2024gpt} and Gemini-2.5-Pro~\cite{comanici2025gemini}, we employ a \textbf{direct prompting} approach, where the models are asked to directly predict a corresponding score. 
We follow the same instruction templates provided in the original dataset papers (listed in Appendix~\ref{appendix:full-prompt-templates}). 
Even under these conditions, the proprietary models still perform worse than AQAScore.
To further validate our method, we evaluate two additional baselines: (1) \textbf{direct prompting} using open-source models, and (2) a \textbf{cascading} approach, which first generates an audio caption and then compares it with the target text using a language-only LLM. 
Results from Qwen2.5-Omni and Audio Flamingo 3 fall behind AQAScore, showing that explicitly modeling AQA probability distributions yields more consistent and human-aligned relevance estimation. 


Among different ALLM backbones, Qwen2.5-Omni generally outperforms Audio Flamingo 3 variants. 
Both models are capable in audio understanding and reasoning, but Qwen2.5-Omni might benefit from substantially larger training data (approximately 3,000k hours vs. 54k hours, as reported in their respective original papers). 
Within Qwen2.5-Omni, the 7B version also surpasses the 3B variant. 
These results suggest that AQAScore is backbone-agnostic and not restricted to a specific architecture; notably, its effectiveness scales with the capability of the underlying ALLM, where stronger models consistently yield higher correlation with human perception. 
Overall, across both human-rated text–audio relevance datasets, AQAScore consistently surpasses all baselines, demonstrating its effectiveness as a perceptually grounded evaluation metric.

\subsection{Pairwise Comparison}

We evaluate pairwise comparison performance across three settings: RELATE-Pair, Baton-Pair, and the original Baton benchmark. 
While RELATE-Pair and Baton-Pair are evaluated using pairwise accuracy, the original \textit{Baton} benchmark is used to report the area under the ROC curve (AUC) under two-event and three-event instruction settings, reflecting different levels of task difficulty. 
As shown in Table~\ref{tab:relate-baton-pairwise}, AQAScore consistently outperforms CLAPScore across all benchmarks and evaluation metrics. On RELATE-Pair, AQAScore achieves the highest pairwise accuracy, with Qwen2.5-Omni-7B reaching 77.6\%, outperforming all CLAPScore variants. Similarly, on Baton-Pair and Baton, AQAScore attains both higher accuracy and superior AUC values, indicating a stronger alignment with human preferences than embedding-based metrics.

We further observe that \textbf{direct prompting} and \textbf{cascading} baselines perform poorly on the Baton benchmark, with AUC values close to 0.5, which corresponds to near-random guessing. This result suggests that directly prompting audio-language models or cascading heuristic judgments (e.g., generating captions before scoring) is insufficient for reliable preference evaluation without an explicit, probability-based formulation. In contrast, the robust performance of AQAScore under multi-event instruction settings supports the effectiveness of reframing evaluation as a probabilistic audio question answering task. Furthermore, consistent with our findings in the relevance task, we observe that AQAScore's performance scales with the capacity of the underlying ALLM, confirming that the framework is backbone-agnostic and benefits from stronger model backbones.

\subsection{Compositional Reasoning Analysis}

To better understand why AQAScore correlates more strongly with human judgment, we evaluate its performance on the CompA benchmark, which measures compositional reasoning across event order (\textit{CompA-order}) and attribute binding (\textit{CompA-attribute}). 
We include \textbf{CompA-CLAP}, the state-of-the-art baseline proposed in the original paper~\cite{ghoshcompa}, which utilizes task-specific data augmentation and training to improve reasoning.

Several key observations emerge from the results in Table~\ref{tab:compa}. 
First, performance in the Text setting (selecting the correct caption given an audio) is generally higher than in the Audio setting (selecting the correct audio given a caption) for all models. 
For example, Qwen2.5-Omni-7B reaches 67.0\% in Order-Text but drops to 52.0\% in Order-Audio. 
This pattern suggests that distinguishing between acoustically similar clips remains more challenging for current models than distinguishing between textual descriptions.
Second, the Attribute task consistently yields lower accuracy than the Order task. 
Correcting binding attributes (e.g., matching a ``laugh'' to a ``baby'' vs. a ``woman'') appears to be a more difficult skill than tracking temporal structure. 
Notably, despite the targeted enhancements of CompA-CLAP, AQAScore using the Qwen2.5-Omni-7B backbone still outperforms this specialized baseline in most settings.

Consistent with our previous findings, these results demonstrate that AQAScore's effectiveness is backbone-agnostic; while it works across different architectures, its reasoning capability scales with the size and capability of the underlying ALLM. 
By leveraging the explicit reasoning of modern ALLMs rather than global embedding similarity, AQAScore provides a more semantically grounded and robust evaluation of complex text–audio alignment.

\subsection{Effect of Question Templates}

To examine the sensitivity of AQAScore to prompt design, we evaluate the model under a set of different question templates, which are detailed in Appendix~\ref{appendix:prompt-template-aqascore-intro}.
To analyze robustness with respect to prompt variation, we report the mean performance and standard deviation across question templates for each benchmark and metric, as shown in Figure~\ref{fig:prompt-varianc}.
Overall, we observe that the choice of question template has a relatively limited impact on performance, indicating that AQAScore is robust to variations in prompt wording.
This result suggests that the model’s judgments are primarily driven by its understanding of audio--text alignment, rather than superficial differences in prompt phrasing.
Full experimental results are provided in Tables~\cref{tab:prompt-sensitive-relate,tab:prompt-sensitive-pam,tab:prompt-sensitive-fense,tab:prompt-sensitive-fense-part2,tab:prompt-sensitive-brace,tab:prompt-sensitive-brace-part2,tab:prompt-sensitive-compa}.
In addition, we examine minor variations in prompt formatting, such as the presence or absence of quotation marks in instructions (e.g., \texttt{Does this audio contain the sound events described by the text: \{description\}?} and \texttt{Does this audio contain the sound events described by the text: ``\{description\}''?}). We find that such formatting differences lead to negligible performance differences. Accordingly, we use the version without quotation marks throughout this paper.

\input{figures/prompt-variance}

%% file: figures/prompt-variance.tex
\begin{figure}[t]
    \centering
    \includegraphics[width=0.48\textwidth]{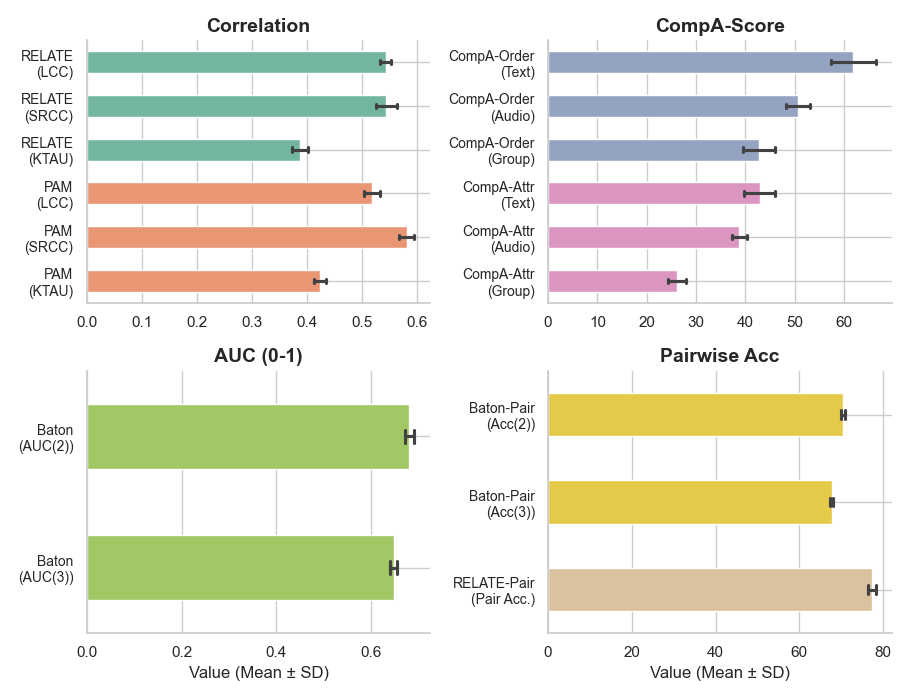}
    \caption{Robustness of AQAScore with Qwen2.5-Omni-7B across diverse question templates. Error bars indicate the standard deviation over prompt variations. 
    The consistently low variance across benchmarks and metrics suggests that the model’s judgments are stable with respect to prompt phrasing.}
    \label{fig:prompt-varianc}
\end{figure}

%% file: sections/discussion.tex

Our results demonstrate that AQAScore is backbone-agnostic and scales effectively with the capability of the underlying ALLM. 
We observe a consistent performance gain as model capacity increases; for example, Qwen2.5-Omni-7B outperforms its 3B variant across both correlation (LCC: 0.544 vs. 0.443) and pairwise accuracy (77.6\% vs. 75.0\%). 
This suggests that the probabilistic verification framework can be generalized to various ALLM architectures, directly benefiting from the continuous evolution of large-scale audio understanding models.
The general superiority of the Qwen family over AF3 variants likely stems from the disparity in their training data scales (approximately 3,000k vs. 54k hours), which provides a more granular acoustic world knowledge for evaluation. 
Specifically, within the AF3 family where different tuning stages are available, we find that AF3-Chat consistently lags behind the Base or Think (reasoning-oriented) versions. 
For instance, on the RELATE benchmark, AF3-Chat achieves a significantly lower LCC (0.3) than AF3-Base (0.475). 
Similarly, on the CompA-Order Text task, AF3-Chat drops to near-random accuracy (24\%) while AF3-Think maintains a much higher 42.25\%.
This degradation may reflect a performance trade-off introduced during the chat-tuning process.
While this process improves conversational helpfulness, it may inadvertently collapse the model’s internal probability distribution. 
For evaluation tasks, such collapsed distributions are less calibrated for the fine-grained log-probability extraction that AQAScore requires. 
In contrast, reasoning-oriented models like AF3-Think better preserve the semantic links between audio and text.

%% file: sections/conclusion.tex
In this work, we present AQAScore, an evaluation framework for assessing text–audio alignment beyond surface-level similarity.
By framing evaluation as targeted semantic verification, AQAScore directly probes whether an audio satisfies specific semantic conditions implied by a text description, rather than relying on global embedding proximity or open-ended generation.
Across diverse evaluation settings, including human-rated relevance, pairwise preference, and compositional reasoning tasks, AQAScore demonstrates consistently stronger alignment with human judgments and higher sensitivity to fine-grained semantic inconsistencies such as temporal order and attribute binding.
Importantly, our results show that AQAScore is backbone-agnostic and scales with the capability of the underlying audio–language model: stronger ALLMs yield more reliable and perceptually grounded evaluation signals.
Overall, this work highlights an alternative perspective on text–audio evaluation: moving from measuring similarity to verifying whether an audio semantically satisfies the content described by the text.

%% file: sections/limitation.tex

\textbf{Evaluation Dimensions and Modalities.}
Our current evaluation primarily focuses on assessing text–audio relevance in general audio generation. Important aspects such as stylistic consistency and the relevance of generated music remain underexplored~\cite{tjandra2025meta, liu2025musiceval}. Future work could extend AQAScore into a multi-dimensional evaluation framework that incorporates these perceptual facets, including emotion, timbre, and musicality, to provide a more holistic assessment of audio generation.
\newline
\newline
\noindent\textbf{Data Scarcity and Human Alignment.} 
The audio generation field lacks extensive human-annotated resources compared to the text-to-image domain. 
Currently, our analysis is constrained by the limited availability of datasets with human-labeled relevance ratings (e.g., RELATE and PAM). 
Expanding this effort to include larger-scale datasets and a wider range of state-of-the-art models remains a priority. 
Given AQAScore’s strong human correlation, a promising future direction is to utilize it as a human-aligned reward model to directly optimize and improve text-to-audio generation systems through reinforcement learning.
\newline
\newline
\noindent\textbf{Benchmark Performance and Model Specialization.} 
While AQAScore demonstrates superior sensitivity to fine-grained semantic nuances and compositional attributes, it does not consistently outperform existing methods in specific text-selection settings, such as the FENSE~\cite{zhou2022can} and BRACE~\cite{guobrace} benchmarks (detailed analysis in Appendix~\ref{appendix:pairwise_caption_fense_brace}). 
This suggests that general-purpose ALLMs may still encounter consistency issues in certain pairwise comparison tasks. 
Future research may investigate developing specialized, evaluation-oriented ALLMs tailored for fine-grained and consistent assessment to further enhance the robustness of audio evaluation metrics.

%% file: sections/ethics_statement.tex
AI-assisted tools were used solely to improve the clarity and fluency of the manuscript.

%% file: sections/appendix.tex
In this section, we present the system prompts and instructions used in all experiments, including the AQAScore approach, prompting setup, and cascade setup.

\subsection{Prompt Templates Used in AQAScore}
\label{appendix:prompt-template-aqascore-intro}
We present the system prompts used in our AQAScore approach in Table~\ref{appendix:system-prompt-aqascore}.
The prompt instructions for RELATE and PAM are shown in Table~\ref{tab:prompt-sensitive-relate} and Table~\ref{tab:prompt-sensitive-pam}, respectively.
For FENSE and BRACE, the prompt instructions are provided in Table~\ref{tab:prompt-sensitive-fense} and Table~\ref{tab:prompt-sensitive-fense-part2}, and Table~\ref{tab:prompt-sensitive-brace} and Table~\ref{tab:prompt-sensitive-brace-part2}, respectively.
For the pairwise settings, Table~\ref{tab:prompt-sensitive-relate} presents the prompt instructions for RELATE-Pair, Table~\ref{tab:prompt-sensitive-baton} for Baton-Pair, and Table~\ref{tab:prompt-sensitive-baton} for Baton.
Finally, the prompt instructions for CompA are shown in Table~\ref{tab:prompt-sensitive-compa}.

\subsection{Prompt Templates Used in Prompting Setup}

In this setup, we directly prompt the model with the audio input, the system prompt, and the instruction, and ask the model to answer according to the instruction.

We provide the system prompts and full prompt templates used in our prompting setups.
For the text–audio relevance experiments, Table~\ref{appendix:system-prompt-prompting-score-relate} shows the system prompts for RELATE and PAM.
Tables~\ref{appendix:prompt-template-prompting-score-relate} and \ref{appendix:prompt-template-pam} present the corresponding prompt instructions for RELATE and PAM, respectively.

For the pairwise comparison experiments,
Table~\ref{appendix:system-prompt-prompting-comparison-relate} shows the system prompts for RELATE-Pair and Baton-Pair,
while Table~\ref{appendix:system-prompt-prompting-fense} shows the system prompts for FENSE and BRACE,
and Table~\ref{appendix:system-prompt-prompting-baton} shows the system prompts for Baton.
Table~\ref{appendix:prompt-template-prompting-relate-pair} presents the prompt instructions for RELATE-Pair and Baton-Pair,
while Table~\ref{appendix:prompt-template-fense} presents the prompt instructions for FENSE and BRACE,
and Table~\ref{appendix:prompt-template-baton} presents the prompt instructions for Baton.
Note that we also explored reasoning in the prompting setups. 
Specifically, we prompted the ALLMs not only for ratings and preferences but also for concise reasoning in their responses. 
However, this did not lead to any noticeable performance improvement.

\subsection{Prompt Templates Used in Cascade Setup}
\label{appendix:full-prompt-templates}

In this setup, we employ a cascade strategy inspired by~\cite{kuan2024speech}.
In the first stage, the ALLM generates a detailed caption for the given audio, using the system prompt shown in Table~\ref{appendix:system-prompt-captioning} and the corresponding prompt instruction presented in Table~\ref{appendix:prompt-template-captioning}.
In the second stage, the ALLM is prompted to answer questions such as whether the audio caption matches the target text and to rate the text–audio relevance on a Likert scale based on the comparison between the generated caption and the target text.
In short, in the second stage, the generated caption is treated as the audio content, and the ALLM must answer the questions solely based on this caption.

We provide the system prompts and full prompt templates used in our cascade setups.
For the text–audio relevance experiments, Table~\ref{appendix:system-prompt-cascade-scoring-relate} shows the system prompts for RELATE and PAM.
Tables~\ref{appendix:prompt-template-cascade-score-relate} and \ref{appendix:prompt-template-cascade-score-pam} present the corresponding prompt instructions for RELATE and PAM, respectively.

For the pairwise comparison experiments,
Table~\ref{appendix:system-prompt-cascade-comparison-relate} provides the system prompt used for RELATE-Pair and Baton-Pair,
Table~\ref{appendix:system-prompt-cascade-baton} provides the system prompt for Baton,
and Table~\ref{appendix:system-prompt-cascade-fense} provides the system prompts for FENSE and BRACE.
The corresponding prompt instructions are shown in
Table~\ref{appendix:prompt-template-cascade-comparison-relate} for RELATE-Pair and Baton-Pair,
Table~\ref{appendix:prompt-template-cascade-baton} for Baton,
and Table~\ref{appendix:prompt-template-cascade-fense} for both FENSE and BRACE.
We also examined the use of reasoning within the cascade setups by prompting the ALLMs to provide concise reasoning in addition to ratings and preferences. 
However, this did not yield any noticeable improvement in performance.

\subsection{CLAPScore Checkpoints}
\label{appendix:clapscore-checkpoints}

To report CLAPScore results, we evaluated several publicly available CLAP pretrained checkpoints~\footnote{\url{huggingface.co/lukewys/laion_clap/tree/main}}. 
Table~\ref{appendix:clap-checkpoints} lists the exact checkpoint identifiers and the corresponding 
training data used for each variant referenced in the main paper.

\subsection{Effect on Question Templates for AQAScore}
\label{appendix:effect-prompt-templates}
We report the results of using different question templates for computing AQAScore in Tables~\cref{tab:prompt-sensitive-relate,tab:prompt-sensitive-pam,tab:prompt-sensitive-fense,tab:prompt-sensitive-fense-part2,tab:prompt-sensitive-brace,tab:prompt-sensitive-brace-part2,tab:prompt-sensitive-compa}.
In addition to the default question templates,
\texttt{``Does this audio contain the sound events described by the text: \{description\}?''}
for RELATE, PAM, RELATE-Pair, and CompA,
and
\texttt{``Does the caption \{description\_1\} better describe the audible sound events in the audio than \{description\_2\}?''}
for FENSE and BRACE,
we further evaluate eight and six alternative question templates, respectively, to examine their impact on model performance.
For brevity, we omit the trailing phrase \texttt{``Please answer yes or no''} from all templates.
The placeholders \texttt{\{description\}}, \texttt{\{description\_1\}}, and \texttt{\{description\_2\}} are replaced with the corresponding target text prompts.

\subsection{Inference Setups}
\label{appendix:inference-setups}
In all experiments, except when calling the APIs of GPT-4o-audio, Gemini-2.5-Flash, and Gemini-2.5-Pro, inference is performed on all models using a single NVIDIA RTX 3090 GPU.
For both the prompting and cascade setups, we use greedy decoding and set the maximum number of new tokens to 512.

\subsection{Pairwise Caption Comparison Benchmarks: FENSE and BRACE}
\label{appendix:pairwise_caption_fense_brace}
\input{appendix/appendix_fense_brace}


\input{prompts/system_prompt}

\input{prompts/system_prompt_relate}
\input{prompts/prompt_templates_relate}
\input{prompts/prompt_template_pam}

\input{prompts/system_prompt_fense}
\input{prompts/prompt_template_fense}

\input{prompts/system_prompt_relate_pair}
\input{prompts/prompt_template_relate_pair}

\input{prompts/system_prompt_baton}
\input{prompts/prompt_template_baton}

\input{prompts/system_prompt_captioning}
\input{prompts/prompt_template_captioning}

\input{prompts/system_prompt_cascade_relate}
\input{prompts/prompt_template_cascade_relate}
\input{prompts/prompt_template_cascade_pam}

\input{prompts/system_prompt_cascade_fense}
\input{prompts/prompt_template_cascade_fense}

\input{prompts/system_prompt_cascade_relate_pair}
\input{prompts/prompt_template_cascade_relate_pair}

\input{prompts/system_prompt_cascade_baton}
\input{prompts/prompt_template_cascade_baton}

\input{tables/clap-checkpoints}

\input{tables/prompt_sensitive_text_relevance}

\input{tables/prompt_sensitive_text_relevance_pam}

\input{tables/prompt_sensitive_fense}

\input{tables/prompt_sensitive_brace}

\input{tables/prompt_sensitive_composition_reasoning}
\input{tables/prompt_sensitive_baton}

%% file: appendix/appendix_fense_brace.tex
In our study, we also evaluate model performance under a pairwise comparison setting using the FENSE and BRACE benchmarks.
These benchmarks provide an audio clip paired with two candidate text captions, and the model is required to select the caption that best matches the audio.

Specifically, FENSE~\cite{zhou2022can} constructs two human-judgment datasets (AudioCaps-Eval and Clotho-Eval) for audio caption evaluation, where annotators are asked to decide which of two captions better describes a given audio clip. 
Following~\cite{zhou2022can}, caption pairs are grouped into four categories: human--human correct (HC), human--human incorrect (HI), human--machine (HM), and machine--machine (MM). 
HC pairs contain two human captions that both correctly describe the same audio, whereas HI pairs also contain two human captions but one caption actually describes a different, randomly selected audio. 
HM pairs are formed by one human caption and one machine-generated caption for the same audio, and MM pairs consist of two machine-generated captions for the same audio.
Although FENSE was originally proposed to benchmark reference-based caption metrics, its pairwise design can be naturally repurposed for audio--text relevance: given an audio clip and two alternative captions, our models are asked which caption better matches the audio, and we measure pairwise accuracy across HC/HI/HM/MM subsets.

BRACE~\cite{guobrace} extends this pairwise comparison paradigm and builds a large-scale, reference-free benchmark for robust audio caption evaluation. 
BRACE consists of two sub-benchmarks: \textit{BRACE-Main} and \textit{BRACE-Hallucination}. 
BRACE-Main is constructed from filtered AudioCaps and Clotho evaluation splits and focuses on fine-grained caption comparison. 
For each audio clip, the authors collect multiple human-written captions and captions generated by recent ALLMs, then form caption pairs and obtain human preference annotations indicating which caption better matches the audio. 
Caption pairs are grouped into three types: HH, HM, and MM, where H and M denote human and machine-generated captions, respectively. 
In our experiments, we follow this setup and ask the model, given an audio clip and a caption pair from one of these types, to choose the caption that is more consistent with the audio, and we report pairwise accuracy on each subset.

BRACE-Hallucination further targets robustness to subtle hallucinations. 
Starting from high-quality audio–caption pairs, one caption in each pair is automatically corrupted by replacing nouns with semantically different but contextually plausible alternatives, and human annotators verify the resulting caption pairs. 
This yields audio–caption pairs where one caption is faithful and the other contains hallucinated content that is not present in the audio. 
We include this track in our evaluation to test whether models can reliably prefer the non-hallucinated caption, providing a complementary view of their robustness to audio–caption misalignment.
While FENSE and BRACE were not originally designed for question-based audio evaluation, their pairwise formulation provides a useful testbed for analyzing relative preference judgments.
We therefore examine AQAScore under these settings to assess whether it aligns with human preferences, and to reveal potential limitations compared to embedding-based metrics and direct ALLM judgments.

\input{tables/fense_benchmark}
\input{tables/brace_benchmark}

To further examine whether AQAScore can also capture relative preferences, we evaluate it on pairwise comparison benchmarks, including FENSE and BRACE. 
Unlike RELATE-Pair, where the model is given two audios and asked to determine which one better matches the text instruction, FENSE and BRACE take the opposite direction: given two text instructions, the model must choose which one better matches the audio content. 
As shown in Table~\ref{tab:fense}, the overall accuracy of AQAScore under Qwen2.5-Omni-7B setup is higher than CLAPScore. 
We also observe that directly prompting ALLMs to select the better-matched text instruction yields the highest accuracy in this task. On the other hand, as shown in Table~\ref{tab:aqascore_BRACE}, AQAScore achieves performance comparable to CLAPScore. However, its performance on the HM subset is notably lower. 
We find that this is due to a mismatch between AQAScore’s judging preference and human judgments. 
Specifically, AQAScore tends to favor perceptually detailed, event-by-event descriptions, whereas human evaluators often prefer concise semantic summaries. 
As a result, AQAScore may assign higher scores to captions that decompose the audio into fine-grained perceptual events, even when humans consider a more abstract description preferable. 
For example, AQAScore prefers `Man speaks, engine starts, then more speech, silence, clicking'' over ``A man speaks and runs a machine'', despite human evaluators favoring the latter. 
Moreover, BRACE provides a hallucination subset designed to test robustness against subtle hallucinations. 
We found that AQAScore achieves over 98\% accuracy on most cases, while CLAPScore remains around 90\%. 
This result indicates that AQAScore can better capture subtle inconsistencies or hallucinations in text descriptions.

%% file: tables/fense_benchmark.tex
\begin{table*}[ht]
\centering
\small
\begin{tabular}{llccccccccccc}
\toprule
\textbf{Methods} & \textbf{Models} & \multicolumn{5}{c}{AudioCaps-Eval} & \multicolumn{5}{c}{Clotho-Eval} \\
\cmidrule(lr){3-7} \cmidrule(lr){8-12}
 &  & HC & HI & HM & MM & All & HC & HI & HM & MM & All \\
\midrule
\multirow{5}{*}{AQAScore} 
& Qwen2.5-Omni-3B 
& 68.3 & 92.4 & 90.8 & 61.8 & 73.8 
& 55.7 & 93.9 & 73.3 & 59.2 & 66.2 \\
& Qwen2.5-Omni-7B 
& 71.6 & 97.3 & 92.7 & \underline{79.8} & \underline{84.1} 
& \textbf{63.8} & 95.5 & 74.1 & 74.2 & \textbf{76.1} \\
& AF3 
& 63.4 & 96.9 & 90.8 & 66.5 & 76.1 
& 63.3 & 91.4 & 72.0 & 70.4 & 73.0 \\
& AF3-Think 
& 64.5 & 96.0 & 90.4 & 73.5 & 79.4 
& 61.4 & 91.8 & 66.8 & 67.8 & 70.6 \\
& AF3-Chat 
& 68.3 & 80.8 & 86.2 & 59.3 & 69.6 
& 61.0 & 79.1 & 65.5 & 64.8 & 66.6 \\
\midrule
\multirow{5}{*}{CLAPScore} 
& Music+AS 
& 63.9 & \underline{98.7} & 91.3 & 70.9 & 78.7 
& 58.6 & 97.1 & 66.8 & 63.8 & 68.8 \\
& 630k-AS 
& 65.0 & 98.2 & 95.4 & 70.3 & 79.3 
& 56.7 & 96.3 & 65.5 & 64.0 & 68.3 \\
& Music+Speech+AS 
& 62.8 & 99.1 & 94.5 & 67.9 & 77.8 
& \underline{63.3} & \textbf{97.5} & 72.4 & 62.3 & 69.5 \\
& Music+Speech 
& 62.8 & \textbf{99.6} & 94.0 & 70.3 & 79.0 
& 59.5 & 95.9 & 71.1 & 66.3 & 70.7 \\
& 630k-best 
& 61.2 & \textbf{99.6} & 94.5 & 73.7 & 80.4 
& 62.4 & 96.3 & 66.0 & 63.0 & 68.6 \\
\midrule
\multirow{5}{*}{Prompting} 
& Qwen2.5-Omni-3B
& \textbf{68.9} & \underline{98.7} & 93.1 & \textbf{81.0} & \textbf{84.6}  
& 61.4 & 92.6 & 73.3 & \underline{72.5} & 74.3 \\
& Qwen2.5-Omni-7B
& 62.3 & 95.1 & 87.6 & 74.0 & 78.6
& 61.9 & 92.2 & 68.1 & \textbf{72.7} & 73.6 \\
& AF3 
& 69.4 & 97.3 & 94.5 & 76.4 & 82.6
& 52.9 & \underline{97.1} & 81.0 & 68.7 & 72.9 \\
& AF3-Think 
& 63.9 & 98.7 & \textbf{97.7} & 76.8 & 82.7
& 58.6 & 80.7 & \underline{76.7} & 66.7 & 69.3 \\
& AF3-Chat 
& 66.1 & 77.7 & 81.7 & 72.4 & 74.1 
& 57.6 & 93.4 & \textbf{77.6} & 71.7 & 74.1 \\
\midrule
\multirow{5}{*}{Cascade} 
& Qwen2.5-Omni-3B
& 48.1 & 97.3 & 89.5 & 71.9 & 76.2
& 55.2 & 95.1 & 70.3 & 68.9 & 71.4 \\
& Qwen2.5-Omni-7B
& 48.6 & 97.3 & 89.0 & 74.0 & 77.2
& 57.1 & 96.7 & 62.5 & 71.2 & 72.0 \\
& AF3 
& 60.7 & 98.2 & \underline{96.3} & 75.2 & 81.1
& 61.4 & 96.7 & 80.6 & 69.6 & \underline{74.4} \\
& AF3-Think 
& 53.6 & 90.6 & 89.5 & 72.3 & 76.0 
& 54.3 & 89.3 & 74.6 & 67.6 & 70.2 \\
& AF3-Chat 
& 60.1 & 77.7 & 75.7 & 61.8 & 67.0
& 55.7 & 60.7 & 48.3 & 55.2 & 55.1 \\
\bottomrule
\end{tabular}
\caption{
Results on the FENSE benchmark for the \textit{AudioCaps-Eval} and \textit{Clotho-Eval} subsets across different models.
Bold indicates the best performance, and underline indicates the second best.
(Unit: \%)
}
\label{tab:fense}
\end{table*}


%% file: tables/brace_benchmark.tex
\begin{table*}[ht]
\centering
\small
\begin{tabular}{llcccccccccc}
\toprule
\textbf{Methods} & \textbf{Models} & \multicolumn{4}{c}{AudioCaps-Main} & \multicolumn{4}{c}{Clotho-Main} & \multicolumn{2}{c}{Hallucination} \\
\cmidrule(lr){3-6} \cmidrule(lr){7-10} \cmidrule(lr){11-12}
 &  & HH & HM & MM & All & HH & HM & MM & All & AudioCaps & Clotho \\
\midrule
\multirow{5}{*}{AQAScore} 
& Qwen2.5-Omni-3B 
& 51.1 & 79.6 & 66.0 & 69.4 
& 68.3 & 72.1 & 67.1 & \underline{69.1} 
& 98.3 & \underline{97.1} \\
& Qwen2.5-Omni-7B 
& 47.5 & 79.8 & 66.8 & 69.5
& 60.3 & 67.9 & 64.5 & 65.2 
& \underline{98.7} & \textbf{98.0} \\
& AF3 
& 46.0 & 64.9 & 60.3 & 60.4
& 64.3 & 61.3 & 58.1 & 60.1 
& 96.9 & 96.9 \\
& AF3-Think 
& 38.9 & 60.0 & 57.8 & 56.3 
& 59.5 & 60.8 & 57.7 & 59.1 
& \textbf{98.8} & \textbf{98.0} \\
& AF3-Chat
& 48.9 & 55.3 & 61.5 & 57.6 
& 60.3 & 56.8 & 59.5 & 58.6 
& 94.6 & 94.5 \\
\midrule
\multirow{5}{*}{CLAPScore} 
& Music+AS 
& 61.9 & \underline{87.6} & 63.5 & 72.7 
& 64.3 & \textbf{73.4} & 62.3 & 66.7 
& 90.8 & 82.0 \\
& 630k-AS 
& 61.7 & 86.3 & 65.1 & \underline{72.8} 
& 62.7 & 69.7 & 63.1 & 65.5 
& 87.6 & 78.0 \\
& Music+Speech+AS 
& 61.9 & 85.6 & 65.1 & 72.7 
& 56.4 & 72.1 & 63.5 & 65.8 
& 90.0 & 81.7 \\
& Music+Speech 
& 59.7 & 80.7 & 66.3 & 71.1 
& 67.5 & 71.1 & 65.1 & 68.0 
& 88.7 & 81.7 \\
& 630k-best 
& 61.9 & \textbf{88.1} & 63.8 & \textbf{73.0} 
& 55.6 & \underline{72.6} & 63.5 & 65.9 
& 91.8 & 82.2 \\
\midrule
\multirow{5}{*}{Prompting} 
& Qwen2.5-Omni-3B 
& 67.6 & 73.0 & \underline{69.2} & 70.5
& 72.2 & 69.0 & \underline{67.9} & 68.8
& 94.9 & 95.4 \\
& Qwen2.5-Omni-7B 
& 58.3 & 73.7 & \textbf{69.7} & 69.9
& \textbf{77.0} & 70.3 & \textbf{70.1} & \textbf{71.0}
& 97.2 & 97.7 \\
& AF3 
& 64.0 & 58.7 & \underline{69.2} & 64.5
& 68.3 & 65.0 & 62.5 & 64.2
& 82.3 & 83.0 \\
& AF3-Think 
& 59.7 & 49.0 & 64.0 & 57.6
& 57.1 & 63.2 & 61.5 & 61.6
& 61.2 & 85.1 \\
& AF3-Chat 
& 70.3 & 63.3 & 65.5 & 66.3
& 53.2 & 63.2 & 60.3 & 60.5
& 87.1 & 85.6 \\
\midrule
\multirow{5}{*}{Cascade} 
& Qwen2.5-Omni-3B 
& 51.8 & 64.5 & 63.3 & 62.4
& 68.3 & 58.7 & 63.9 & 62.5
& 91.2 & 90.2 \\
& Qwen2.5-Omni-7B 
& 57.6 & 69.4 & 62.9 & 64.8
& \underline{76.2} & 64.2 & 65.7 & 66.4
& 95.4 & 95.9 \\
& AF3 
& 56.8 & 52.6 & 67.9 & 60.6
& 69.1 & 62.9 & 65.7 & 65.0
& 87.5 & 85.0 \\
& AF3-Think 
& 52.5 & 49.9 & 57.6 & 54.0
& 62.7 & 59.0 & 62.9 & 61.4
& 67.5 & 74.2 \\
& AF3-Chat 
& 54.7 & 58.2 & 57.8 & 57.6
& 53.2 & 51.1 & 56.5 & 54.0
& 90.5 & 85.1 \\
\bottomrule
\end{tabular}
\caption{
Results on the BRACE benchmark for the \textit{AudioCaps-Main}, \textit{Clotho-Main}, and \textit{Hallucination} subsets across different models. Bold indicates the best performance, and underline indicates the second best. 
(Unit: \%).
}
\label{tab:aqascore_BRACE}
\end{table*}

%% file: prompts/system_prompt.tex
\begin{table}[tbp]
    \footnotesize
    \centering
    \caption{
    System prompt used in the AQAScore setup.
    }
    \begin{tabular}{p{23em}}
    \hline
    \texttt{Your role is to listen attentively to the given audio and decide whether the provided text accurately and completely describes what is heard.} \\
    \texttt{Make your judgment strictly based on the sounds in the audio --- do not guess, imagine, or add information that is not clearly audible.} \\
    \texttt{If something is missing, unclear, or uncertain, do not assume it exists.} \\
    \texttt{} \\
    \texttt{Your task: given an audio clip and a text description, carefully compare the text with what you actually hear.} \\
    \texttt{Identify the main sound events (such as speech, background noise, music, or environmental sounds) and decide whether the text correctly reflects them.} \\
    \texttt{Always respond objectively and concisely, using ``yes'' or ``no'' to indicate whether the text matches the audio content.} \\
    \hline
    \end{tabular}
    \label{appendix:system-prompt-aqascore}
\end{table}

%% file: prompts/system_prompt_relate.tex
\begin{table}[tbp]
    \footnotesize
    \centering
    \caption{
    System prompt used in the prompting setup for RELATE~\cite{kanamori2025relate} and PAM~\cite{deshmukh2024pam}. 
    }
    \begin{tabular}{p{23em}}
    \hline
    \texttt{You are an unbiased and careful audio relevance judge.} \\
    \texttt{Your task is to listen attentively to the audio and assess how well it matches the provided text description.} \\
    \texttt{Base your judgment strictly on what is clearly audible — do not guess, imagine, or add information that is not supported by the sound.} \\
    \texttt{Focus on the actual audible events, their characteristics, and whether they align with the text.} \\
    \texttt{If something is unclear, missing, or uncertain in the audio, do not assume it exists.} \\
    \texttt{Follow the scoring scale provided, and output only the required fields.} \\
    \hline
    \end{tabular}
    \label{appendix:system-prompt-prompting-score-relate}
\end{table}

%% file: prompts/prompt_templates_relate.tex
\begin{table}[tbp]
    \footnotesize
    \centering
    \caption{
    Full prompt template used in the prompting setup for RELATE~\cite{kanamori2025relate}. 
    The placeholder \{text\} is replaced with the actual text caption.
    The scoring scale is defined based on the original datasets and follows the specifications in the original paper.
    }
    \begin{tabular}{p{23em}}
    \hline
    \texttt{How well does this audio match with the provided description?} \\
    \texttt{Description: \{text\}} \\
    \\
    \texttt{Rate the match from 0 to 10 using the following scale:} \\
    \texttt{0 --- Does not match at all} \\
    \texttt{2 --- Has significant discrepancies} \\
    \texttt{5 --- Has several minor discrepancies} \\
    \texttt{8 --- Has a few minor discrepancies} \\
    \texttt{10 --- Matched exactly} \\
    \texttt{} \\
    \texttt{Output your answer in exactly this format:} \\
    \texttt{Score: <number>} \\
    \hline
    \end{tabular}
    \label{appendix:prompt-template-prompting-score-relate}
\end{table}

%% file: prompts/prompt_template_pam.tex
\begin{table}[tbp]
    \footnotesize
    \centering
    \caption{
    Full prompt template used in the prompting setup for PAM~\cite{deshmukh2024pam}. 
    The placeholder \{text\} is replaced with the actual text caption.
    The scoring scale is defined based on the original datasets and follows the specifications in the original paper.
    }
    \begin{tabular}{p{23em}}
    \hline
    \texttt{How well does this audio match with the provided description?} \\
    \texttt{Description: \{text\}} \\
    \\
    \texttt{Rate the match from 1 to 5 using the following scale:} \\
    \texttt{1 (Poor) Audio has minimal or no relevance to the text.} \\
    \texttt{2 (Fair) Audio shows limited relevance to the text.} \\
    \texttt{3 (Good) Audio is adequately relevant to the text.} \\
    \texttt{4 (Very Good) Audio is highly relevant to the text.} \\
    \texttt{5 (Excellent) Audio perfectly matches the text.} \\
    \texttt{} \\
    \texttt{Output your answer in exactly this format:} \\
    \texttt{Score: <number>} \\
    \hline
    \end{tabular}
    \label{appendix:prompt-template-pam}
\end{table}

%% file: prompts/system_prompt_fense.tex
\begin{table}[tbp]
    \footnotesize
    \centering
    \caption{
    System prompt used in the prompting setup for FENSE~\cite{zhou2022can} and BRACE~\cite{guobrace}. 
    }
    \begin{tabular}{p{23em}}
    \hline
    \texttt{You are an unbiased and careful audio-caption matching judge.} \\
    \texttt{Your task is to listen attentively to the audio and compare how well the two captions describe what is actually heard.} \\
    \texttt{Base your judgment strictly on audible evidence — do not guess, infer hidden causes, or add information that is not supported by the sound.} \\
    \texttt{Evaluate which caption aligns more accurately with the audible events, their characteristics, and their temporal structure.} \\
    \texttt{If a detail is unclear or not audible, treat it as not present.} \\
    \texttt{You must choose exactly one caption as the better match and provide a brief, factual reason before giving your final choice.} \\
    \hline
    \end{tabular}
    \label{appendix:system-prompt-prompting-fense}
\end{table}

%% file: prompts/prompt_template_fense.tex
\begin{table}[tbp]
    \footnotesize
    \centering
    \caption{
    Full prompt template used in the prompting setup for FENSE~\cite{zhou2022can} and BRACE~\cite{guobrace}.
    The placeholders \{audio\_caption1\} and \{audio\_caption2\} are replaced with the actual text captions.
    }
    \begin{tabular}{p{23em}}
    \hline
    \texttt{You are given two independently written captions for the same audio clip.} \\
    \\
    \texttt{- Caption 1:} \\
    \\
    \texttt{\{audio\_caption1\}}
    \\ \\
    \texttt{- Caption 2:} \\
    \\
    \texttt{\{audio\_caption2\}}
    \\ \\
    \texttt{Listen to the audio and decide which caption better matches what is heard.} \\
    \texttt{You must choose exactly one option — either 1 or 2. Do not output ties, uncertainty, or alternative formats.} 
    \\ \\
    \texttt{Output your choice in this exact format:} \\
    \texttt{Better Caption: <1 or 2>} \\
    \hline
    \end{tabular}
    \label{appendix:prompt-template-fense}
\end{table}

%% file: prompts/system_prompt_relate_pair.tex
\begin{table}[tbp]
    \footnotesize
    \centering
    \caption{
    System prompt used in the prompting setup for RELATE-Pair~\cite{kanamori2025relate} and Baton-Pair~\cite{liao2024baton}. 
    }
    \begin{tabular}{p{23em}}
    \hline
    \texttt{You are an unbiased and careful audio-text matching judge.} \\
    \texttt{Your task is to compare two audio clips and determine which one more accurately matches the given text description.} \\
    \texttt{Base your judgment strictly on what is actually heard in the audio — do not guess, infer hidden context, or add information that is not present.} \\
    \texttt{Evaluate which audio aligns more closely in content, detail accuracy, and overall relevance to the description.} \\
    \texttt{If a sound is unclear or missing, do not assume it exists.} \\
    \texttt{You must choose exactly one audio clip as the better match.} \\
    \hline
    \end{tabular}
    \label{appendix:system-prompt-prompting-comparison-relate}
\end{table}

%% file: prompts/prompt_template_relate_pair.tex
\begin{table}[tbp]
    \footnotesize
    \centering
    \caption{
    Full prompt template used in the prompting setup for RELATE-Pair~\cite{kanamori2025relate} and Baton-Pair~\cite{liao2024baton}.
    The placeholder \{text\} is replaced with the actual caption.
    }
    \begin{tabular}{p{23em}}
    \hline
    \texttt{Given the text description: \{text\}} 
    \\ \\
    \texttt{Which audio matches the description more accurately?} 
    \\ \\
    \texttt{Output your answer in exactly this format:} \\
    \texttt{Better Match: <first audio or second audio>} \\
    \hline
    \end{tabular}
    \label{appendix:prompt-template-prompting-relate-pair}
\end{table}


%% file: prompts/system_prompt_baton.tex
\begin{table}[tbp]
    \footnotesize
    \centering
    \caption{
    System prompt used in the prompting setup for Baton~\cite{liao2024baton}. 
    }
    \begin{tabular}{p{23em}}
    \hline
    \texttt{You are an unbiased and careful audio relevance judge.} \\
    \texttt{Your task is to listen attentively to the audio and assess how well it matches the provided text description.} \\
    \texttt{Base your judgment strictly on what is clearly audible — do not guess, imagine, or add information that is not supported by the sound.} \\
    \texttt{Focus on the actual audible events, their characteristics, and whether they align with the text.} \\
    \texttt{If something is unclear, missing, or uncertain in the audio, do not assume it exists.} \\
    \texttt{Do not generate any text other than the final decision. Follow the decision rule provided, and output only the required field.} \\
    \hline
    \end{tabular}
    \label{appendix:system-prompt-prompting-baton}
\end{table}

%% file: prompts/prompt_template_baton.tex
\begin{table}[tbp]
    \footnotesize
    \centering
    \caption{
    Full prompt template used in the prompting setup for Baton~\cite{liao2024baton}. 
    The placeholder \{text\} is replaced with the actual text caption.
    }
    \begin{tabular}{p{23em}}
    \hline
    \texttt{How well does this audio match with the provided description?} \\
    \texttt{Description: \{text\}} \\
    \\
    \texttt{Based on the audio content, decide whether this audio should be accepted or rejected for the given description.} \\
    \\
    \texttt{Decision rule:} \\
    \texttt{- Preferred — The audio clearly matches the description and includes all described sound events.} \\
    \texttt{- Rejected — One or more described sound events are missing, unclear, or incorrect.} \\
    \texttt{Output your answer in exactly this format:} \\
    \texttt{Answer: <preferred or rejected>} \\
    \hline
    \end{tabular}
    \label{appendix:prompt-template-baton}
\end{table}

%% file: prompts/system_prompt_captioning.tex
\begin{table}[tbp]
    \footnotesize
    \centering
    \caption{
    System prompt used for generating audio captions in the cascade setup.
    }
    \begin{tabular}{p{23em}}
    \hline
    \texttt{You are a precise and unbiased audio captioning agent.} \\
    \texttt{Your job is to listen carefully to the audio and generate an accurate, detailed caption.} \\
    \texttt{Describe only what is clearly audible — never infer hidden causes, intentions, or off-audio events.} \\
    \texttt{Include all audible sound events, their acoustic characteristics, background sounds, and temporal order when perceivable.} \\
    \texttt{If something is unclear or uncertain, do not guess or fill in details.} \\
    \hline
    \end{tabular}
    \label{appendix:system-prompt-captioning}
\end{table}

%% file: prompts/prompt_template_captioning.tex
\begin{table}[tbp]
    \footnotesize
    \centering
    \caption{
    Full prompt template used for generating audio captions in the cascade setup.
    }
    \begin{tabular}{p{23em}}
    \hline
    \texttt{Listen to the audio and generate a detailed caption.} \\
    \texttt{Describe only what is clearly audible — do not add assumptions or hidden causes.} \\
    \texttt{Include all audible sound events, their sound characteristics, background sounds, and the temporal order of events when perceivable.} 
    \\ \\
    \texttt{Output only the caption:.} 
    \\
    \hline
    \end{tabular}
    \label{appendix:prompt-template-captioning}
\end{table}

%% file: prompts/system_prompt_cascade_relate.tex
\begin{table}[tbp]
    \footnotesize
    \centering
    \caption{
    System prompt used in the cascade setup for RELATE~\cite{kanamori2025relate} and PAM~\cite{deshmukh2024pam}.
    }
    \begin{tabular}{p{23em}}
    \hline
    \texttt{You are an unbiased and careful text-matching evaluator.} \\
    \texttt{Your task is to assess how well the audio caption aligns with the ground-truth description.} \\
    \texttt{Make your judgment strictly based on the textual content—do not guess, add information, or infer anything beyond what is explicitly written.} \\
    \texttt{Evaluate the alignment in terms of meaning, coverage, and detail accuracy.} \\
    \texttt{Follow the scoring scale provided, and output only the required fields.} \\
    \hline
    \end{tabular}
    \label{appendix:system-prompt-cascade-scoring-relate}
\end{table}

%% file: prompts/prompt_template_cascade_relate.tex
\begin{table}[tbp]
    \footnotesize
    \centering
    \caption{
    Full prompt template used in the cascade setup for RELATE~\cite{kanamori2025relate}. 
    The placeholder \{ground\_truth\_description\} is replaced with the actual text caption and \{audio\_caption\} is replaced with the text caption generated by the ALLM.
    The scoring scale is defined based on the original paper.
    }
    \begin{tabular}{p{23em}}
    \hline
    \texttt{Evaluate how well the audio caption matches the ground-truth description.} 
    \\ \\
    \texttt{- Ground-truth description:}
    \\ \\
    \texttt{\{ground\_truth\_description\}}
    \\ \\
    \texttt{- Audio caption:}
    \\ \\
    \texttt{\{audio\_caption\}}
    \\ \\
    \texttt{Rate the match from 0 to 10 using the following scale:} \\
    \texttt{0 --- Does not match at all} \\
    \texttt{2 --- Has significant discrepancies} \\
    \texttt{5 --- Has several minor discrepancies} \\
    \texttt{8 --- Has a few minor discrepancies} \\
    \texttt{10 --- Matched exactly} \\
    \texttt{} \\
    \texttt{Output your answer in exactly this format:} \\
    \texttt{Score: <number>} \\
    \hline
    \end{tabular}
    \label{appendix:prompt-template-cascade-score-relate}
\end{table}

%% file: prompts/prompt_template_cascade_pam.tex
\begin{table}[tbp]
    \footnotesize
    \centering
    \caption{
    Full prompt template used in the cascade setup for PAM~\cite{deshmukh2024pam}. 
    The placeholder \{ground\_truth\_description\} is replaced with the actual text caption and \{audio\_caption\} is replaced with the text caption generated by the ALLM.
    The scoring scale is defined based on the original paper.
    }
    \begin{tabular}{p{23em}}
    \hline
    \texttt{Evaluate how well the audio caption matches the ground-truth description.} 
    \\ \\
    \texttt{- Ground-truth description:}
    \\ \\
    \texttt{\{ground\_truth\_description\}}
    \\ \\
    \texttt{- Audio caption:}
    \\ \\
    \texttt{\{audio\_caption\}}
    \\ \\
    \texttt{Rate the match from 1 to 5 using the following scale:} \\
    \texttt{1 (Poor) Audio has minimal or no relevance to the text.} \\
    \texttt{2 (Fair) Audio shows limited relevance to the text.} \\
    \texttt{3 (Good) Audio is adequately relevant to the text.} \\
    \texttt{4 (Very Good) Audio is highly relevant to the text.} \\
    \texttt{5 (Excellent) Audio perfectly matches the text.} \\
    \texttt{} \\
    \texttt{Output your answer in exactly this format:} \\
    \texttt{Score: <number>} \\
    \hline
    \end{tabular}
    \label{appendix:prompt-template-cascade-score-pam}
\end{table}

%% file: prompts/system_prompt_cascade_fense.tex
\begin{table}[tbp]
    \footnotesize
    \centering
    \caption{
    System prompt used in the cascade setup for FENSE~\cite{zhou2022can} and BRACE~\cite{guobrace}. 
    }
    \begin{tabular}{p{23em}}
    \hline
    \texttt{You are an unbiased and careful text-description matching judge.} \\
    \texttt{Your task is to compare two audio captions and determine which one more accurately matches the ground-truth description.} \\
    \texttt{Base your judgment strictly on textual content — do not guess, infer hidden causes, or add information that is not stated.} \\
    \texttt{Evaluate which caption aligns more closely in meaning, detail accuracy, and coverage of the ground-truth description.} \\
    \texttt{If something is unclear or missing in a caption, treat it as not present.} \\
    \texttt{You must choose exactly one caption as the better match and provide a brief, factual reason before giving your final choice.} \\
    \hline
    \end{tabular}
    \label{appendix:system-prompt-cascade-fense}
\end{table}

%% file: prompts/prompt_template_cascade_fense.tex
\begin{table}[tbp]
    \footnotesize
    \centering
    \caption{
    Full prompt template used in the cascade setup for FENSE~\cite{zhou2022can} and BRACE~\cite{guobrace}. 
    The placeholder \{ground\_truth\_description\} is replaced with the actual text caption, and \{audio\_caption1\} and \{audio\_caption2\} are replaced with the text captions generated by the ALLM.
    }
    \begin{tabular}{p{23em}}
    \hline
    \texttt{Evaluate which audio caption better matches the ground-truth description.} 
    \\ \\
    \texttt{- Ground-truth description:}
    \\ \\
    \texttt{\{ground\_truth\_description\}}
    \\ \\
    \texttt{- Audio caption 1:}
    \\ \\
    \texttt{\{audio\_caption1\}}
    \\ \\
    \texttt{- Audio caption 2:}
    \\ \\
    \texttt{\{audio\_caption2\}}
    \\ \\
    \texttt{Decide which caption aligns more closely with the ground-truth description.} \\
    \texttt{You must choose exactly one option — either 1 or 2. Do not output ties, uncertainty, or alternative formats.} 
    \\ \\
    \texttt{Give your final choice in exactly this format:} \\
    \texttt{Better Caption: <1 or 2>} \\
    \hline
    \end{tabular}
    \label{appendix:prompt-template-cascade-fense}
\end{table}

%% file: prompts/system_prompt_cascade_relate_pair.tex
\begin{table}[tbp]
    \footnotesize
    \centering
    \caption{
    System prompt used in the cascade setup for RELATE-Pair~\cite{kanamori2025relate} and Baton-Pair~\cite{liao2024baton}. 
    }
    \begin{tabular}{p{23em}}
    \hline
    \texttt{You are an unbiased and careful text-description matching judge.} \\
    \texttt{Your task is to compare two candidate captions and determine which one more accurately matches the ground-truth description.} \\
    \texttt{Base your judgment strictly on the textual content — do not guess, imagine, or add information that is not stated.} \\
    \texttt{Evaluate which caption aligns more closely in meaning, detail accuracy, and coverage.} \\
    \texttt{If something is unclear or missing in a caption, do not assume it exists.} \\
    \texttt{You must choose exactly one caption as the better match.} \\
    \hline
    \end{tabular}
    \label{appendix:system-prompt-cascade-comparison-relate}
\end{table}

%% file: prompts/prompt_template_cascade_relate_pair.tex
\begin{table}[tbp]
    \footnotesize
    \centering
    \caption{
    Full prompt template used in the cascade setup for RELATE-Pair~\cite{kanamori2025relate} and Baton-Pair~\cite{liao2024baton}. 
    The placeholder \{ground\_truth\_description\} is replaced with the actual text caption, and \{audio\_caption1\} and \{audio\_caption2\} are replaced with the text captions generated by the ALLM.
    }
    \begin{tabular}{p{23em}}
    \hline
    \texttt{Evaluate which audio caption better matches the ground-truth description.} 
    \\ \\
    \texttt{- Ground-truth description:}
    \\ \\
    \texttt{\{ground\_truth\_description\}}
    \\ \\
    \texttt{- Audio caption 1:}
    \\ \\
    \texttt{\{audio\_caption1\}}
    \\ \\
    \texttt{- Audio caption 2:}
    \\ \\
    \texttt{\{audio\_caption2\}}
    \\ \\
    \texttt{Decide which caption is a closer match to the ground-truth description.} \\
    \texttt{You must choose exactly one caption — either 1 or 2. Do not output ties, uncertainty, or additional options.} 
    \\ \\
    \texttt{Output your choice in this exact format:} \\
    \texttt{Better Caption: <1 or 2>} \\
    \hline
    \end{tabular}
    \label{appendix:prompt-template-cascade-comparison-relate}
\end{table}

%% file: prompts/system_prompt_cascade_baton.tex
\begin{table}[tbp]
    \footnotesize
    \centering
    \caption{
    System prompt used in the cascade setup for Baton~\cite{liao2024baton}. 
    }
    \begin{tabular}{p{23em}}
    \hline
    \texttt{You are an unbiased and careful text-matching evaluator.} \\
    \texttt{Your task is to assess whether the audio caption aligns with the ground-truth description.} \\
    \texttt{Make your judgment strictly based on the textual content—do not guess, add information, or infer anything beyond what is explicitly written.} \\
    \texttt{Evaluate the alignment in terms of meaning, coverage, and detail accuracy.} \\
    \texttt{Follow the decision rule provided, and output only the required field.} \\
    \hline
    \end{tabular}
    \label{appendix:system-prompt-cascade-baton}
\end{table}

%% file: prompts/prompt_template_cascade_baton.tex
\begin{table}[tbp]
    \footnotesize
    \centering
    \caption{
    Full prompt template used in the cascade setup for Baton~\cite{liao2024baton}. 
    The placeholder \{ground\_truth\_description\} is replaced with the actual text caption and \{audio\_caption\} is replaced with the text caption generated by the ALLM.
    }
    \begin{tabular}{p{23em}}
    \hline
    \texttt{Evaluate whether the audio caption matches the ground-truth description.} 
    \\ \\
    \texttt{- Ground-truth description:}
    \\ \\
    \texttt{\{ground\_truth\_description\}}
    \\ \\
    \texttt{- Audio caption:}
    \\ \\
    \texttt{\{audio\_caption\}}
    \\ \\
    \texttt{Decision rule:} \\
    \texttt{- Preferred — The audio caption accurately covers the meaning of the ground-truth description, including all key sound events.} \\
    \texttt{- Rejected — The audio caption misses important sound events, contains incorrect information, or does not faithfully reflect the description.} \\
    \texttt{Output your answer in exactly this format:} \\
    \texttt{Decision: <preferred or rejected>} \\
    \hline
    \end{tabular}
    \label{appendix:prompt-template-cascade-baton}
\end{table}

%% file: tables/clap-checkpoints.tex
\begin{table*}[tbp]
\centering
\small
\caption{Summary of CLAPScore checkpoints used in our experiments.}
\label{tab:clapscore-checkpoints}
\begin{tabular}{lll}
\toprule
\textbf{Name in Paper} & \textbf{Checkpoint Identifier} & \textbf{Training Data} \\
\midrule
Music+AS            
& \texttt{music\_audioset\_epoch\_15\_esc\_90.14.pt}
& Music + Audioset + LAION-Audio-630k
\\
630k-AS             
& \texttt{630k-audioset-best.pt}           
& LAION-Audio-630k + AudioSet 
\\
Music+Speech+AS     
& \texttt{music\_speech\_audioset\_epoch\_15\_esc\_89.98.pt}    
& Music + Speech + Audioset + LAION-Audio-630k
\\
Music+Speech        
& \texttt{music\_speech\_epoch\_15\_esc\_89.25.pt}              
& Music + Speech + LAION-Audio-630k
\\
630k-best           
& \texttt{clap-630k-best.pt}                  
& LAION-Audio-630k
\\
\bottomrule
\end{tabular}
\label{appendix:clap-checkpoints}
\end{table*}

%% file: tables/prompt_sensitive_text_relevance.tex
\begin{table*}[ht]
\centering
\small
\begin{tabular}{l cccc}
\toprule
{Prompt Templates} 
& LCC $\uparrow$ & SRCC $\uparrow$ & KTAU $\uparrow$ & Pair Acc. $\uparrow$ \\
\midrule
\multicolumn{5}{l}{{AQAScore (Qwen2.5-Omni-3B)}} \\
\textit{Does the audio contain the sound events described by the text: \{text\}?} 
& 0.443 & 0.453 & 0.327 & {75.00} \\
\textit{Is this audio clip containing the sound events described by the text: \{text\}?} 
& 0.459 & 0.457 & 0.330 & 74.29 \\
\textit{Does this recording include the sound events described by the text: \{text\}?} 
& {0.461} & {0.466} & {0.336} & 74.29 \\
\textit{Does this audio include the sound events described by the text: \{text\}?} 
& 0.457 & 0.463 & 0.335 & 74.05 \\
\textit{Does the audio contain the sound events mentioned in the text: \{text\}?} 
& 0.438 & 0.448 & 0.323 & 74.76 \\
\textit{Are the sound events described by the text: \{text\} heard in the audio?} 
& {0.464} & 0.460 & 0.334 & 73.33 \\
\textit{Are the sound events described in the text: \{text\} present in this audio?} 
& 0.453 & 0.452 & 0.327 & 74.29 \\
\textit{Can the sound events described by the text: \{text\} be heard in this audio?} 
& 0.452 & 0.443 & 0.321 & 73.81 \\
\textit{Do the sound events described by the text: \{text\} occur in the audio?} 
& 0.443 & 0.431 & 0.313 & 73.57 \\
\midrule
\multicolumn{5}{l}{{AQAScore (Qwen2.5-Omni-7B)}} \\
\textit{Does the audio contain the sound events described by the text: \{text\}?} 
& 0.544 & 0.556 & 0.396 & 77.62 \\
\textit{Is this audio clip containing the sound events described by the text: \{text\}?} 
& 0.548 & 0.552 & 0.393 & 76.90 \\
\textit{Does this recording include the sound events described by the text: \{text\}?} 
& {0.558} & {0.565} & {0.404} & 77.86 \\
\textit{Does this audio include the sound events described by the text: \{text\}?} 
& {0.556} & {0.565} & {0.404} & {78.10} \\
\textit{Does the audio contain the sound events mentioned in the text: \{text\}?} 
& 0.540 & 0.556 & 0.396 & 77.86 \\
\textit{Are the sound events described by the text: \{text\} heard in the audio?} 
& 0.529 & 0.522 & 0.370 & 76.67 \\
\textit{Are the sound events described in the text: \{text\} present in this audio?} 
& 0.540 & 0.534 & 0.380 & 75.71 \\
\textit{Can the sound events described by the text: \{text\} be heard in this audio?} 
& 0.533 & 0.516 & 0.367 & 75.48 \\
\textit{Do the sound events described by the text: \{text\} occur in the audio?} 
& 0.533 & 0.525 & 0.373 & 75.71 \\
\midrule
\multicolumn{5}{l}{{AQAScore (AF3)}} \\
\textit{Does the audio contain the sound events described by the text: \{text\}?} 
& 0.475 & 0.508 & 0.357 & 73.57 \\
\textit{Is this audio clip containing the sound events described by the text: \{text\}?} 
& {0.485} & {0.517} & {0.364} & {76.43} \\
\textit{Does this recording include the sound events described by the text: \{text\}?} 
& {0.490} & 0.503 & 0.353 & 73.33 \\
\textit{Does this audio include the sound events described by the text: \{text\}?} 
& 0.475 & 0.501 & 0.352 & 71.67 \\
\textit{Does the audio contain the sound events mentioned in the text: \{text\}?} 
& 0.413 & 0.506 & 0.356 & 73.81 \\
\textit{Are the sound events described by the text: \{text\} heard in the audio?} 
& 0.464 & 0.466 & 0.325 & 73.33 \\
\textit{Are the sound events described in the text: \{text\} present in this audio?} 
& 0.489 & 0.500 & 0.350 & {76.43} \\
\textit{Can the sound events described by the text: \{text\} be heard in this audio?} 
& 0.474 & 0.465 & 0.327 & 75.48 \\
\textit{Do the sound events described by the text: \{text\} occur in the audio?} 
& 0.460 & 0.479 & 0.334 & 75.48 \\
\midrule
\multicolumn{5}{l}{{AQAScore (AF3-Think)}} \\
\textit{Does the audio contain the sound events described by the text: \{text\}?} 
& 0.435 & {0.474} & {0.330} & 72.38 \\
\textit{Is this audio clip containing the sound events described by the text: \{text\}?} 
& 0.365 & 0.410 & 0.282 & 67.62 \\
\textit{Does this recording include the sound events described by the text: \{text\}?} 
& 0.418 & 0.456 & 0.317 & 70.00 \\
\textit{Does this audio include the sound events described by the text: \{text\}?} 
& 0.438 & 0.468 & 0.326 & 71.90 \\
\textit{Does the audio contain the sound events mentioned in the text: \{text\}?} 
& {0.461} & {0.475} & {0.330} & 71.43 \\
\textit{Are the sound events described by the text: \{text\} heard in the audio?} 
& 0.397 & 0.433 & 0.300 & 66.90 \\
\textit{Are the sound events described in the text: \{text\} present in this audio?} 
& 0.387 & 0.400 & 0.275 & 67.38 \\
\textit{Can the sound events described by the text: \{text\} be heard in this audio?} 
& 0.386 & 0.393 & 0.270 & 68.10 \\
\textit{Do the sound events described by the text: \{text\} occur in the audio?} 
& {0.450} & 0.460 & 0.318 & {71.67} \\
\midrule
\multicolumn{5}{l}{{AQAScore (AF3-Chat)}} \\
\textit{Does the audio contain the sound events described by the text: \{text\}?} 
& 0.300 & 0.389 & 0.271 & {72.14} \\
\textit{Is this audio clip containing the sound events described by the text: \{text\}?} 
& 0.327 & 0.404 & 0.282 & 70.24 \\
\textit{Does this recording include the sound events described by the text: \{text\}?} 
& {0.345} & 0.410 & 0.285 & 71.43 \\
\textit{Does this audio include the sound events described by the text: \{text\}?} 
& 0.327 & {0.417} & {0.291} & 71.67 \\
\textit{Does the audio contain the sound events mentioned in the text: \{text\}?} 
& 0.292 & 0.401 & 0.280 & 69.76 \\
\textit{Are the sound events described by the text: \{text\} heard in the audio?} 
& 0.276 & 0.340 & 0.232 & 65.00 \\
\textit{Are the sound events described in the text: \{text\} present in this audio?} 
& 0.331 & 0.377 & 0.261 & 64.05 \\
\textit{Can the sound events described by the text: \{text\} be heard in this audio?} 
& 0.272 & 0.306 & 0.213 & 67.62 \\
\textit{Do the sound events described by the text: \{text\} occur in the audio?} 
& 0.265 & {0.390} & {0.272} & {68.33} \\
\bottomrule
\end{tabular}
\caption{
Prompt sensitivity analysis for AQAScore variants on the RELATE benchmark.
The placeholder \texttt{\{text\}} is replaced with the corresponding target text prompt.
}
\label{tab:prompt-sensitive-relate}
\end{table*}

%% file: tables/prompt_sensitive_text_relevance_pam.tex
\begin{table*}[ht]
\centering
\small
\begin{tabular}{lccc}
\toprule
{Prompt Templates}
& LCC $\uparrow$ & SRCC $\uparrow$ & KTAU $\uparrow$ \\
\midrule
\multicolumn{4}{l}{{AQAScore (Qwen2.5-Omni-3B)}} \\
\textit{Does the audio contain the sound events described by the text: \{text\}?} 
& 0.540 & 0.560 & 0.410 \\
\textit{Is this audio clip containing the sound events described by the text: \{text\}?} 
& 0.546 & 0.556 & 0.405 \\
\textit{Does this recording include the sound events described by the text: \{text\}?} 
& 0.547 & 0.564 & {0.412} \\
\textit{Does this audio include the sound events described by the text: \{text\}?} 
& 0.548 & {0.566} & 0.414 \\
\textit{Does the audio contain the sound events mentioned in the text: \{text\}?} 
& 0.526 & 0.557 & 0.407 \\
\textit{Are the sound events described by the text: \{text\} heard in the audio?} 
& {0.556} & {0.571} & {0.419} \\
\textit{Are the sound events described in the text: \{text\} present in this audio?} 
& 0.539 & 0.538 & 0.395 \\
\textit{Can the sound events described by the text: \{text\} be heard in this audio?} 
& 0.551 & 0.558 & 0.413 \\
\textit{Do the sound events described by the text: \{text\} occur in the audio?} 
& 0.529 & 0.551 & 0.405 \\
\midrule
\multicolumn{4}{l}{{AQAScore (Qwen2.5-Omni-7B)}} \\
\textit{Does the audio contain the sound events described by the text: \{text\}?} 
& 0.518 & 0.589 & 0.429 \\
\textit{Is this audio clip containing the sound events described by the text: \{text\}?} 
& 0.522 & 0.582 & 0.423 \\
\textit{Does this recording include the sound events described by the text: \{text\}?} 
& {0.531} & {0.602} & {0.442} \\
\textit{Does this audio include the sound events described by the text: \{text\}?} 
& 0.526 & 0.589 & 0.430 \\
\textit{Does the audio contain the sound events mentioned in the text: \{text\}?} 
& 0.522 & 0.586 & 0.427 \\
\textit{Are the sound events described by the text: \{text\} heard in the audio?} 
& {0.532} & 0.571 & 0.414 \\
\textit{Are the sound events described in the text: \{text\} present in this audio?} 
& 0.514 & 0.556 & 0.403 \\
\textit{Can the sound events described by the text: \{text\} be heard in this audio?} 
& 0.487 & 0.581 & 0.425 \\
\textit{Do the sound events described by the text: \{text\} occur in the audio?} 
& 0.505 & 0.569 & 0.416 \\
\midrule
\multicolumn{4}{l}{{AQAScore (AF3)}} \\
\textit{Does the audio contain the sound events described by the text: \{text\}?} 
& 0.496 & 0.538 & 0.383 \\
\textit{Is this audio clip containing the sound events described by the text: \{text\}?} 
& 0.524 & 0.549 & 0.391 \\
\textit{Does this recording include the sound events described by the text: \{text\}?} 
& 0.527 & 0.551 & 0.393 \\
\textit{Does this audio include the sound events described by the text: \{text\}?} 
& 0.507 & 0.540 & 0.385 \\
\textit{Does the audio contain the sound events mentioned in the text: \{text\}?} 
& 0.486 & 0.565 & 0.403 \\
\textit{Are the sound events described by the text: \{text\} heard in the audio?} 
& 0.543 & 0.554 & 0.394 \\
\textit{Are the sound events described in the text: \{text\} present in this audio?} 
& {0.571} & {0.587} & {0.420} \\
\textit{Can the sound events described by the text: \{text\} be heard in this audio?} 
& 0.542 & 0.548 & 0.390 \\
\textit{Do the sound events described by the text: \{text\} occur in the audio?} 
& {0.553} & {0.583} & {0.417} \\
\midrule
\multicolumn{4}{l}{{AQAScore (AF3-Think)}} \\
\textit{Does the audio contain the sound events described by the text: \{text\}?} 
& 0.582 & 0.587 & 0.419 \\
\textit{Is this audio clip containing the sound events described by the text: \{text\}?} 
& 0.498 & 0.542 & 0.387 \\
\textit{Does this recording include the sound events described by the text: \{text\}?} 
& 0.587 & 0.601 & 0.431 \\
\textit{Does this audio include the sound events described by the text: \{text\}?} 
& 0.603 & 0.608 & 0.437 \\
\textit{Does the audio contain the sound events mentioned in the text: \{text\}?} 
& 0.600 & 0.597 & 0.428 \\
\textit{Are the sound events described by the text: \{text\} heard in the audio?} 
& 0.562 & 0.591 & 0.428 \\
\textit{Are the sound events described in the text: \{text\} present in this audio?} 
& 0.564 & 0.552 & 0.395 \\
\textit{Can the sound events described by the text: \{text\} be heard in this audio?} 
& 0.600 & 0.591 & 0.429 \\
\textit{Do the sound events described by the text: \{text\} occur in the audio?} 
& {0.636} & {0.635} & {0.461} \\
\midrule
\multicolumn{4}{l}{{AQAScore (AF3-Chat)}} \\
\textit{Does the audio contain the sound events described by the text: \{text\}?} 
& 0.381 & 0.435 & 0.310 \\
\textit{Is this audio clip containing the sound events described by the text: \{text\}?} 
& 0.354 & 0.404 & 0.285 \\
\textit{Does this recording include the sound events described by the text: \{text\}?} 
& 0.418 & 0.442 & 0.313 \\
\textit{Does this audio include the sound events described by the text: \{text\}?} 
& 0.400 & 0.442 & 0.314 \\
\textit{Does the audio contain the sound events mentioned in the text: \{text\}?} 
& {0.405} & {0.477} & {0.340} \\
\textit{Are the sound events described by the text: \{text\} heard in the audio?} 
& 0.375 & 0.427 & 0.305 \\
\textit{Are the sound events described in the text: \{text\} present in this audio?} 
& {0.413} & {0.460} & {0.326} \\
\textit{Can the sound events described by the text: \{text\} be heard in this audio?} 
& 0.249 & 0.297 & 0.207 \\
\textit{Do the sound events described by the text: \{text\} occur in the audio?} 
& 0.310 & 0.408 & 0.290 \\
\bottomrule
\end{tabular}
\caption{
Prompt sensitivity analysis for AQAScore variants on the PAM benchmark.
The placeholder \texttt{\{text\}} is replaced with the corresponding target text prompt.
}
\label{tab:prompt-sensitive-pam}
\end{table*}

%% file: tables/prompt_sensitive_fense.tex
\newcommand{\pairprompt}[1]{\textit{#1}}

\begin{table*}[ht]
\centering
\small
\setlength{\tabcolsep}{3pt}
\renewcommand{\arraystretch}{1.15}

\begin{tabularx}{\textwidth}{%
>{\raggedright\arraybackslash}X
ccccccccccc}
\toprule
& \multicolumn{5}{c}{\textbf{AudioCaps-Eval}} & & \multicolumn{5}{c}{\textbf{Clotho-Eval}} \\
\cmidrule(lr){2-6} \cmidrule(lr){8-12}
\textbf{Prompt Templates} & HC & HI & HM & MM & All & & HC & HI & HM & MM & All \\
\midrule

\multicolumn{12}{l}{\textbf{AQAScore (Qwen2.5-Omni-3B)}} \\

\pairprompt{Does the caption \{text1\} better describe the audible sound events in the audio than \{text2\}?}
& 68.3 & 92.4 & 90.8 & 61.8 & 73.8
&& 55.7 & 93.9 & 73.3 & 59.2 & 66.2 \\

\pairprompt{Is the caption \{text1\} a more accurate description of the audible sound events in the audio than \{text2\}?}
& 67.8 & 92.9 & 89.0 & 55.5 & 70.5
&& 55.7 & 93.4 & 67.2 & 57.0 & 64.1 \\

\pairprompt{Does \{text1\} describe the actual audible sound events in the audio more accurately than \{text2\}?}
& 66.1 & 63.0 & 74.8 & 44.0 & 56.5
&& 48.6 & 77.9 & 60.8 & 47.5 & 54.4 \\

\pairprompt{Is \{text1\} more faithful to the audible sound events in the audio than \{text2\}?}
& 66.1 & 90.6 & 92.2 & 54.8 & 70.0
&& 55.7 & 93.4 & 69.0 & 55.9 & 63.7 \\

\pairprompt{Compared to \{text2\}, does \{text1\} provide a more accurate description of the audible sound events in the audio?}
& 65.0 & 98.2 & 92.2 & 83.4 & 85.0
&& 62.4 & 95.1 & 72.0 & 73.9 & 75.4 \\

\pairprompt{Does \{text1\} better reflect the sound events that are actually audible in the audio than \{text2\}?}
& 65.0 & 80.8 & 88.5 & 57.4 & 68.6
&& 54.3 & 85.3 & 71.6 & 58.0 & 63.8 \\

\midrule
\multicolumn{12}{l}{\textbf{AQAScore (Qwen2.5-Omni-7B)}} \\

\pairprompt{Does the caption \{text1\} better describe the audible sound events in the audio than \{text2\}?}
& 71.6 & 97.3 & 92.7 & 79.8 & 84.1
&& 63.8 & 95.5 & 74.1 & 74.2 & 76.1 \\

\pairprompt{Is the caption \{text1\} a more accurate description of the audible sound events in the audio than \{text2\}?}
& 75.4 & 96.4 & 92.7 & 81.5 & 85.4
&& 63.3 & 94.3 & 75.0 & 73.8 & 75.8 \\

\pairprompt{Does \{text1\} describe the actual audible sound events in the audio more accurately than \{text2\}?}
& 72.1 & 96.4 & 92.7 & 81.0 & 84.6
&& 63.3 & 95.1 & 73.3 & 74.1 & 75.8 \\

\pairprompt{Is \{text1\} more faithful to the audible sound events in the audio than \{text2\}?}
& 71.6 & 96.0 & 92.7 & 78.0 & 83.1
&& 62.4 & 95.1 & 73.7 & 72.8 & 75.1 \\

\pairprompt{Compared to \{text2\}, does \{text1\} provide a more accurate description of the audible sound events in the audio?}
& 71.0 & 96.9 & 92.7 & 83.1 & 85.6
&& 62.4 & 94.7 & 72.0 & 73.5 & 75.1 \\

\pairprompt{Does \{text1\} better reflect the sound events that are actually audible in the audio than \{text2\}?}
& 70.5 & 96.9 & 92.7 & 72.6 & 80.5
&& 63.8 & 95.1 & 73.3 & 73.4 & 75.5 \\

\bottomrule
\end{tabularx}

\caption{
Prompt sensitivity analysis for pairwise AQAScore variants on the FENSE benchmark.
The placeholders \texttt{\{text1\}} and \texttt{\{text2\}} are replaced with the corresponding target text prompts. (Unit: \%).
}
\label{tab:prompt-sensitive-fense}
\end{table*}

\begin{table*}[ht]
\centering
\small
\setlength{\tabcolsep}{3pt}
\renewcommand{\arraystretch}{1.15}

\begin{tabularx}{\textwidth}{%
>{\raggedright\arraybackslash}X
ccccccccccc}
\toprule
& \multicolumn{5}{c}{\textbf{AudioCaps-Eval}} & & \multicolumn{5}{c}{\textbf{Clotho-Eval}} \\
\cmidrule(lr){2-6} \cmidrule(lr){8-12}
\textbf{Prompt Templates} & HC & HI & HM & MM & All & & HC & HI & HM & MM & All \\
\midrule

\multicolumn{12}{l}{\textbf{AQAScore (AF3)}} \\

\pairprompt{Does the caption \{text1\} better describe the audible sound events in the audio than \{text2\}?}
& 63.4 & 96.9 & 90.8 & 66.5 & 76.1
&& 63.3 & 91.4 & 72.0 & 70.4 & 73.0 \\

\pairprompt{Is the caption \{text1\} a more accurate description of the audible sound events in the audio than \{text2\}?}
& 65.6 & 97.8 & 93.1 & 60.7 & 74.3
&& 64.3 & 91.8 & 72.8 & 66.6 & 71.2 \\

\pairprompt{Does \{text1\} describe the actual audible sound events in the audio more accurately than \{text2\}?}
& 51.9 & 97.3 & 90.4 & 57.8 & 70.2
&& 62.9 & 90.2 & 65.1 & 63.5 & 67.9 \\

\pairprompt{Is \{text1\} more faithful to the audible sound events in the audio than \{text2\}?}
& 56.8 & 96.9 & 90.4 & 67.0 & 75.3
&& 61.4 & 88.5 & 67.2 & 68.5 & 70.5 \\

\pairprompt{Compared to \{text2\}, does \{text1\} provide a more accurate description of the audible sound events in the audio?}
& 57.4 & 92.9 & 92.2 & 69.1 & 76.0
&& 62.4 & 76.2 & 67.2 & 62.0 & 65.1 \\

\pairprompt{Does \{text1\} better reflect the sound events that are actually audible in the audio than \{text2\}?}
& 62.3 & 96.4 & 91.7 & 69.6 & 77.6
&& 64.3 & 90.6 & 69.0 & 69.7 & 72.2 \\

\midrule
\multicolumn{12}{l}{\textbf{AQAScore (AF3-Think)}} \\

\pairprompt{Does the caption \{text1\} better describe the audible sound events in the audio than \{text2\}?}
& 64.5 & 96.0 & 90.4 & 73.5 & 79.4
&& 61.4 & 91.8 & 66.8 & 67.8 & 70.6 \\

\pairprompt{Is the caption \{text1\} a more accurate description of the audible sound events in the audio than \{text2\}?}
& 70.0 & 94.2 & 91.7 & 60.9 & 74.1
&& 60.5 & 88.9 & 67.2 & 60.3 & 65.9 \\

\pairprompt{Does \{text1\} describe the actual audible sound events in the audio more accurately than \{text2\}?}
& 66.1 & 92.9 & 90.4 & 60.7 & 73.0
&& 59.1 & 86.5 & 62.1 & 56.3 & 62.3 \\

\pairprompt{Is \{text1\} more faithful to the audible sound events in the audio than \{text2\}?}
& 65.0 & 94.2 & 88.1 & 72.3 & 78.1
&& 64.3 & 86.5 & 63.8 & 61.0 & 65.9 \\

\pairprompt{Compared to \{text2\}, does \{text1\} provide a more accurate description of the audible sound events in the audio?}
& 65.6 & 94.2 & 91.7 & 71.6 & 78.6
&& 63.3 & 83.6 & 65.1 & 64.2 & 67.3 \\

\pairprompt{Does \{text1\} better reflect the sound events that are actually audible in the audio than \{text2\}?}
& 67.8 & 96.0 & 92.7 & 77.8 & 82.4
&& 63.8 & 87.3 & 60.8 & 59.7 & 64.8 \\

\midrule
\multicolumn{12}{l}{\textbf{AQAScore (AF3-Chat)}} \\

\pairprompt{Does the caption \{text1\} better describe the audible sound events in the audio than \{text2\}?}
& 68.3 & 80.8 & 86.2 & 59.3 & 69.6
&& 61.0 & 79.1 & 65.5 & 64.8 & 66.6 \\

\pairprompt{Is the caption \{text1\} a more accurate description of the audible sound events in the audio than \{text2\}?}
& 67.2 & 79.0 & 82.1 & 55.0 & 66.3
&& 61.0 & 82.0 & 65.1 & 61.1 & 65.0 \\

\pairprompt{Does \{text1\} describe the actual audible sound events in the audio more accurately than \{text2\}?}
& 69.4 & 80.8 & 86.7 & 52.4 & 66.5
&& 60.0 & 78.7 & 69.0 & 61.5 & 65.1 \\

\pairprompt{Is \{text1\} more faithful to the audible sound events in the audio than \{text2\}?}
& 70.0 & 75.0 & 83.9 & 58.5 & 68.0
&& 55.2 & 76.6 & 62.9 & 63.9 & 64.6 \\

\pairprompt{Compared to \{text2\}, does \{text1\} provide a more accurate description of the audible sound events in the audio?}
& 56.8 & 75.0 & 83.0 & 72.8 & 72.6
&& 59.5 & 62.3 & 60.3 & 62.6 & 61.8 \\

\pairprompt{Does \{text1\} better reflect the sound events that are actually audible in the audio than \{text2\}?}
& 68.9 & 76.3 & 89.5 & 60.9 & 70.2
&& 60.0 & 78.3 & 64.2 & 63.1 & 65.2 \\

\bottomrule
\end{tabularx}

\caption{
Prompt sensitivity analysis for pairwise AQAScore variants on the FENSE benchmark.
The placeholders \texttt{\{text1\}} and \texttt{\{text2\}} are replaced with the corresponding target text prompts. (Unit: \%).
}
\label{tab:prompt-sensitive-fense-part2}
\end{table*}

%% file: tables/prompt_sensitive_brace.tex
\begin{table*}[ht]
\centering
\small
\setlength{\tabcolsep}{4pt}
\renewcommand{\arraystretch}{1.15}

\begin{tabularx}{\textwidth}{
>{\raggedright\arraybackslash}X
cccc
cccc
cc}
\toprule
& \multicolumn{4}{c}{\textbf{AudioCaps-Main}} &
\multicolumn{4}{c}{\textbf{Clotho-Main}} &
\multicolumn{2}{c}{\textbf{Hallu.}} \\
\cmidrule(lr){2-5} \cmidrule(lr){6-9} \cmidrule(lr){10-11}
\textbf{Prompt Templates}
& HH & HM & MM & All
& HH & HM & MM & All
& AC & CL \\
\midrule

\multicolumn{11}{l}{\textbf{AQAScore (Qwen2.5-Omni-3B)}} \\

\pairprompt{Does the caption \{text1\} better describe the audible sound events than the caption \{text2\}?}
& 66.2 & 80.2 & 70.8 & 73.9 & 65.9 & 69.7 & 73.5 & 71.1 & 98.3 & 98.1 \\

\pairprompt{Is the caption \{text1\} a more accurate description of the audible sound events than the caption \{text2\}?}
& 63.3 & 56.2 & 60.3 & 59.0 & 67.5 & 65.3 & 69.7 & 67.7 & 31.8 & 67.1 \\

\pairprompt{Does the caption \{text1\} describe the audible sound events more accurately than the caption \{text2\}?}
& 64.0 & 53.0 & 59.9 & 57.7 & 63.5 & 65.8 & 67.5 & 66.3 & 38.6 & 65.9 \\

\pairprompt{Is the caption \{text1\} more faithful to the audible sound events than the caption \{text2\}?}
& 58.3 & 55.1 & 58.8 & 57.3 & 62.7 & 60.5 & 66.1 & 63.6 & 16.1 & 50.7 \\

\pairprompt{Compared to the caption \{text2\}, does the caption \{text1\} provide a more accurate description of the audible sound events?}
& 61.2 & 65.2 & 62.2 & 63.2 & 58.7 & 62.4 & 66.7 & 64.1 & 37.0 & 60.3 \\

\pairprompt{Does the caption \{text1\} better reflect the sound events that are audible than the caption \{text2\}?}
& 61.2 & 62.5 & 60.6 & 61.4 & 61.1 & 60.5 & 65.9 & 63.3 & 34.2 & 59.8 \\

\pairprompt{Does the caption \{text1\} provide a more accurate and fluent description of the audio than the caption \{text2\}?}
& 62.6 & 64.5 & 61.3 & 62.7 & 61.1 & 66.3 & 68.5 & 66.7 & 43.4 & 73.9 \\

\midrule
\multicolumn{11}{l}{\textbf{AQAScore (Qwen2.5-Omni-7B)}} \\

\pairprompt{Does the caption \{text1\} better describe the audible sound events than the caption \{text2\}?}
& 61.9 & 73.7 & 68.8 & 69.9 & 72.2 & 70.5 & 72.7 & 71.8 & 99.1 & 99.2 \\

\pairprompt{Is the caption \{text1\} a more accurate description of the audible sound events than the caption \{text2\}?}
& 60.4 & 73.3 & 67.6 & 68.9 & 65.1 & 68.4 & 72.7 & 70.1 & 98.5 & 97.6 \\

\pairprompt{Does the caption \{text1\} describe the audible sound events more accurately than the caption \{text2\}?}
& 63.3 & 70.3 & 67.4 & 68.0 & 65.9 & 70.3 & 70.7 & 69.9 & 98.0 & 96.7 \\

\pairprompt{Is the caption \{text1\} more faithful to the audible sound events than the caption \{text2\}?}
& 62.6 & 72.1 & 66.1 & 68.0 & 66.7 & 65.3 & 69.5 & 67.5 & 98.6 & 98.0 \\

\pairprompt{Compared to the caption \{text2\}, does the caption \{text1\} provide a more accurate description of the audible sound events?}
& 59.7 & 67.4 & 67.4 & 66.5 & 66.7 & 64.2 & 69.7 & 67.2 & 98.3 & 97.1 \\

\pairprompt{Does the caption \{text1\} better reflect the sound events that are audible than the caption \{text2\}?}
& 59.7 & 73.0 & 66.8 & 68.4 & 66.7 & 65.5 & 68.7 & 67.2 & 98.5 & 97.8 \\

\pairprompt{Does the caption \{text1\} provide a more accurate and fluent description of the audio than the caption \{text2\}?}
& 59.0 & 73.0 & 70.2 & 70.0 & 66.7 & 67.1 & 72.7 & 69.8 & 98.0 & 96.6 \\

\bottomrule
\end{tabularx}

\caption{
Prompt sensitivity analysis for AQAScore variants on the BRACE benchmark
under pairwise comparison.
AC: AudioCaps. CL: Clotho.
The placeholders \texttt{\{text1\}} and \texttt{\{text2\}} are replaced with the corresponding target text prompts. (Unit: \%).
}
\label{tab:prompt-sensitive-brace}
\end{table*}

\begin{table*}[ht]
\centering
\small
\setlength{\tabcolsep}{4pt}
\renewcommand{\arraystretch}{1.15}

\begin{tabularx}{\textwidth}{
>{\raggedright\arraybackslash}X
cccc
cccc
cc}
\toprule
& \multicolumn{4}{c}{\textbf{AudioCaps-Main}} &
\multicolumn{4}{c}{\textbf{Clotho-Main}} &
\multicolumn{2}{c}{\textbf{Hallu.}} \\
\cmidrule(lr){2-5} \cmidrule(lr){6-9} \cmidrule(lr){10-11}
\textbf{Prompt Templates}
& HH & HM & MM & All
& HH & HM & MM & All
& AC & CL \\
\midrule

\multicolumn{11}{l}{\textbf{AQAScore (AF3)}} \\

\pairprompt{Does the caption \{text1\} better describe the audible sound events than the caption \{text2\}?}
& 64.8 & 76.4 & 59.2 & 66.6
& 66.7 & 68.2 & 66.1 & 66.9
& 99.0 & 97.4 \\

\pairprompt{Is the caption \{text1\} a more accurate description of the audible sound events than the caption \{text2\}?}
& 64.0 & 71.5 & 64.9 & 67.3
& 65.9 & 66.1 & 67.9 & 66.9
& 98.8 & 98.3 \\

\pairprompt{Does the caption \{text1\} describe the audible sound events more accurately than the caption \{text2\}?}
& 67.6 & 76.2 & 64.4 & 69.3
& 73.8 & 65.8 & 68.9 & 68.3
& 99.1 & 98.4 \\

\pairprompt{Is the caption \{text1\} more faithful to the audible sound events than the caption \{text2\}?}
& 61.2 & 65.6 & 64.0 & 64.3
& 73.0 & 66.1 & 68.5 & 68.1
& 98.6 & 97.2 \\

\pairprompt{Compared to the caption \{text2\}, does the caption \{text1\} provide a more accurate description of the audible sound events?}
& 69.1 & 76.4 & 63.3 & 69.1
& 68.3 & 69.2 & 67.5 & 68.2
& 97.8 & 96.6 \\

\pairprompt{Does the caption \{text1\} better reflect the sound events that are audible than the caption \{text2\}?}
& 63.3 & 72.8 & 63.6 & 67.2
& 65.1 & 67.1 & 66.7 & 66.6
& 98.8 & 97.8 \\

\pairprompt{Does the caption \{text1\} provide a more accurate and fluent description of the audio than the caption \{text2\}?}
& 67.6 & 72.6 & 68.3 & 69.9
& 66.7 & 66.3 & 70.1 & 68.2
& 98.7 & 97.6 \\

\midrule
\multicolumn{11}{l}{\textbf{AQAScore (AF3-Think)}} \\

\pairprompt{Does the caption \{text1\} better describe the audible sound events than the caption \{text2\}?}
& 67.6 & 70.8 & 64.7 & 67.4
& 61.1 & 60.0 & 67.9 & 64.1
& 96.7 & 94.6 \\

\pairprompt{Is the caption \{text1\} a more accurate description of the audible sound events than the caption \{text2\}?}
& 61.2 & 63.6 & 57.6 & 60.4
& 65.9 & 52.9 & 60.1 & 58.1
& 96.7 & 95.5 \\

\pairprompt{Does the caption \{text1\} describe the audible sound events more accurately than the caption \{text2\}?}
& 62.6 & 64.9 & 59.2 & 61.8
& 65.9 & 54.0 & 61.7 & 59.3
& 96.7 & 95.3 \\

\pairprompt{Is the caption \{text1\} more faithful to the audible sound events than the caption \{text2\}?}
& 56.8 & 68.3 & 57.9 & 61.8
& 67.5 & 55.5 & 64.3 & 61.4
& 97.1 & 92.6 \\

\pairprompt{Compared to the caption \{text2\}, does the caption \{text1\} provide a more accurate description of the audible sound events?}
& 59.0 & 63.6 & 59.0 & 60.8
& 66.7 & 52.9 & 58.9 & 57.6
& 96.9 & 93.8 \\

\pairprompt{Does the caption \{text1\} better reflect the sound events that are audible than the caption \{text2\}?}
& 55.4 & 64.3 & 54.2 & 58.3
& 65.9 & 52.4 & 59.5 & 57.6
& 97.9 & 95.5 \\

\pairprompt{Does the caption \{text1\} provide a more accurate and fluent description of the audio than the caption \{text2\}?}
& 71.2 & 68.5 & 63.3 & 66.3
& 62.7 & 57.6 & 64.1 & 61.5
& 96.0 & 93.5 \\

\midrule
\multicolumn{11}{l}{\textbf{AQAScore (AF3-Chat)}} \\

\pairprompt{Does the caption \{text1\} better describe the audible sound events than the caption \{text2\}?}
& 52.5 & 64.3 & 63.1 & 62.3
& 65.1 & 57.1 & 58.9 & 59.0
& 98.4 & 98.2 \\

\pairprompt{Is the caption \{text1\} a more accurate description of the audible sound events than the caption \{text2\}?}
& 66.2 & 74.2 & 65.1 & 68.7
& 69.1 & 65.8 & 65.5 & 66.0
& 91.6 & 88.0 \\

\pairprompt{Does the caption \{text1\} describe the audible sound events more accurately than the caption \{text2\}?}
& 69.1 & 78.2 & 63.6 & 70.0
& 69.8 & 66.1 & 66.3 & 66.6
& 97.0 & 95.5 \\

\pairprompt{Is the caption \{text1\} more faithful to the audible sound events than the caption \{text2\}?}
& 61.2 & 67.0 & 64.5 & 65.1
& 71.4 & 63.2 & 69.1 & 67.1
& 94.0 & 89.2 \\

\pairprompt{Compared to the caption \{text2\}, does the caption \{text1\} provide a more accurate description of the audible sound events?}
& 62.6 & 76.9 & 63.1 & 68.4
& 72.2 & 62.9 & 64.3 & 64.8
& 91.9 & 87.1 \\

\pairprompt{Does the caption \{text1\} better reflect the sound events that are audible than the caption \{text2\}?}
& 65.5 & 72.8 & 61.7 & 66.5
& 68.3 & 61.6 & 62.9 & 63.1
& 89.7 & 87.4 \\

\pairprompt{Does the caption \{text1\} provide a more accurate and fluent description of the audio than the caption \{text2\}?}
& 68.4 & 74.6 & 64.7 & 69.0
& 70.6 & 63.7 & 63.1 & 64.3
& 97.6 & 94.6 \\

\bottomrule
\end{tabularx}

\caption{
Prompt sensitivity analysis for AQAScore variants on the BRACE benchmark
under pairwise comparison.
AC: AudioCaps. CL: Clotho.
The placeholders \texttt{\{text1\}} and \texttt{\{text2\}} are replaced with the corresponding target text prompts. (Unit: \%).
}
\label{tab:prompt-sensitive-brace-part2}
\end{table*}

%% file: tables/prompt_sensitive_composition_reasoning.tex
\newcommand{\singleprompt}[1]{\textit{#1}}

\begin{table*}[ht]
\centering
\small
\setlength{\tabcolsep}{4pt}
\renewcommand{\arraystretch}{1.15}

\begin{tabularx}{\textwidth}{%
>{\raggedright\arraybackslash}X
ccc
ccc}
\toprule
\textbf{Prompt Templates}
& \multicolumn{3}{c}{\textbf{CompA-Order}}
& \multicolumn{3}{c}{\textbf{CompA-Attribute}} \\
\cmidrule(lr){2-4} \cmidrule(lr){5-7}
& Text & Audio & Group & Text & Audio & Group \\
\midrule

\multicolumn{7}{l}{\textbf{AQAScore (Qwen2.5-Omni-3B)}} \\

\singleprompt{Does the audio contain the sound events described by \{text\}?}
& 50.00 & 35.00 & 25.25 & 29.95 & 23.35 & 14.21 \\

\singleprompt{Is this audio clip containing the sound events described by \{text\}?}
& 49.25 & 34.00 & 25.25 & 28.43 & 23.35 & 14.72 \\

\singleprompt{Does this recording include the sound events described by \{text\}?}
& 49.75 & 32.50 & 24.50 & 27.41 & 20.30 & 11.68 \\

\singleprompt{Does this audio include the sound events described by \{text\}?}
& 50.00 & 35.00 & 26.00 & 28.93 & 22.34 & 13.71 \\

\singleprompt{Does the audio contain the sound events mentioned in \{text\}?}
& 47.50 & 34.75 & 24.75 & 31.47 & 22.34 & 13.71 \\

\singleprompt{Are the sound events described by \{text\} heard in the audio?}
& 48.50 & 33.50 & 26.50 & 24.37 & 22.84 & 12.69 \\

\singleprompt{Are the sound events described in \{text\} present in this audio?}
& 46.00 & 31.50 & 23.75 & 28.43 & 23.35 & 13.20 \\

\singleprompt{Can the sound events described by \{text\} be heard in this audio?}
& 34.75 & 26.75 & 16.75 & 26.40 & 18.27 & 13.71 \\

\singleprompt{Do the sound events described by \{text\} occur in the audio?}
& 45.75 & 33.25 & 24.25 & 27.92 & 22.34 & 13.20 \\

\midrule
\multicolumn{7}{l}{\textbf{AQAScore (Qwen2.5-Omni-7B)}} \\

\singleprompt{Does the audio contain the sound events described by \{text\}?}
& 67.00 & 52.00 & 45.75 & 44.16 & 37.06 & 24.87 \\

\singleprompt{Is this audio clip containing the sound events described by \{text\}?}
& 67.50 & 52.25 & 45.50 & 44.16 & 35.53 & 24.87 \\

\singleprompt{Does this recording include the sound events described by \{text\}?}
& 66.00 & 50.00 & 43.75 & 43.65 & 34.52 & 23.35 \\

\singleprompt{Does this audio include the sound events described by \{text\}?}
& 65.75 & 50.75 & 44.50 & 44.16 & 35.03 & 23.35 \\

\singleprompt{Does the audio contain the sound events mentioned in \{text\}?}
& 65.75 & 52.25 & 45.25 & 44.16 & 39.59 & 26.40 \\

\singleprompt{Are the sound events described by \{text\} heard in the audio?}
& 64.50 & 51.50 & 42.75 & 43.15 & 34.01 & 25.38 \\

\singleprompt{Are the sound events described in \{text\} present in this audio?}
& 64.25 & 51.50 & 43.25 & 39.59 & 36.55 & 23.86 \\

\singleprompt{Can the sound events described by \{text\} be heard in this audio?}
& 56.75 & 46.00 & 35.50 & 42.64 & 35.03 & 24.87 \\

\singleprompt{Do the sound events described by \{text\} occur in the audio?}
& 63.50 & 50.25 & 43.00 & 41.12 & 31.98 & 22.34 \\

\midrule
\multicolumn{7}{l}{\textbf{AQAScore (AF3)}} \\

\singleprompt{Does the audio contain the sound events described by \{text\}?}
& 51.00 & 42.50 & 32.25 & 41.33 & 33.67 & 18.88 \\

\singleprompt{Is this audio clip containing the sound events described by \{text\}?}
& 49.75 & 38.25 & 28.25 & 43.88 & 33.16 & 18.88 \\

\singleprompt{Does this recording include the sound events described by \{text\}?}
& 51.50 & 38.00 & 29.25 & 42.86 & 29.59 & 17.86 \\

\singleprompt{Does this audio include the sound events described by \{text\}?}
& 50.00 & 40.00 & 29.75 & 43.88 & 29.59 & 17.86 \\

\singleprompt{Does the audio contain the sound events mentioned in \{text\}?}
& 54.50 & 42.50 & 33.00 & 41.33 & 31.63 & 17.35 \\

\singleprompt{Are the sound events described by \{text\} heard in the audio?}
& 49.50 & 38.75 & 31.25 & 40.31 & 33.67 & 20.92 \\

\singleprompt{Are the sound events described in \{text\} present in this audio?}
& 49.75 & 39.50 & 30.50 & 41.84 & 32.14 & 19.39 \\

\singleprompt{Can the sound events described by \{text\} be heard in this audio?}
& 44.25 & 34.75 & 24.00 & 36.73 & 26.53 & 15.82 \\

\singleprompt{Do the sound events described by \{text\} occur in the audio?}
& 49.50 & 36.00 & 27.00 & 38.78 & 34.18 & 18.88 \\

\midrule
\multicolumn{7}{l}{\textbf{AQAScore (AF3-Think)}} \\

\singleprompt{Does the audio contain the sound events described by \{text\}?}
& 42.25 & 27.75 & 17.00 & 33.67 & 22.45 & 11.73 \\

\singleprompt{Is this audio clip containing the sound events described by \{text\}?}
& 41.00 & 26.00 & 16.25 & 37.24 & 13.78 & 8.16 \\

\singleprompt{Does this recording include the sound events described by \{text\}?}
& 39.50 & 24.75 & 15.75 & 32.65 & 20.92 & 11.73 \\

\singleprompt{Does this audio include the sound events described by \{text\}?}
& 39.50 & 26.75 & 16.75 & 31.63 & 21.43 & 11.73 \\

\singleprompt{Does the audio contain the sound events mentioned in \{text\}?}
& 45.75 & 28.00 & 18.25 & 36.22 & 20.41 & 11.73 \\

\singleprompt{Are the sound events described by \{text\} heard in the audio?}
& 40.50 & 26.25 & 19.00 & 35.20 & 20.92 & 11.22 \\

\singleprompt{Are the sound events described in \{text\} present in this audio?}
& 38.75 & 27.50 & 16.25 & 31.12 & 23.47 & 11.73 \\

\singleprompt{Can the sound events described by \{text\} be heard in this audio?}
& 33.50 & 21.50 & 11.25 & 29.59 & 21.94 & 13.27 \\

\singleprompt{Do the sound events described by \{text\} occur in the audio?}
& 39.50 & 24.25 & 15.50 & 36.73 & 24.49 & 15.31 \\

\midrule
\multicolumn{7}{l}{\textbf{AQAScore (AF3-Chat)}} \\

\singleprompt{Does the audio contain the sound events described by \{text\}?}
& 24.00 & 28.00 & 14.00 & 17.86 & 14.80 & 4.08 \\

\singleprompt{Is this audio clip containing the sound events described by \{text\}?}
& 24.25 & 26.50 & 12.50 & 22.45 & 19.90 & 7.14 \\

\singleprompt{Does this recording include the sound events described by \{text\}?}
& 24.75 & 27.00 & 12.00 & 20.41 & 17.86 & 6.12 \\

\singleprompt{Does this audio include the sound events described by \{text\}?}
& 26.25 & 28.00 & 13.50 & 20.92 & 19.39 & 7.65 \\

\singleprompt{Does the audio contain the sound events mentioned in \{text\}?}
& 25.75 & 27.00 & 14.75 & 19.39 & 17.86 & 4.59 \\

\singleprompt{Are the sound events described by \{text\} heard in the audio?}
& 25.75 & 26.25 & 13.50 & 22.45 & 21.43 & 10.71 \\

\singleprompt{Are the sound events described in \{text\} present in this audio?}
& 23.00 & 27.25 & 12.50 & 23.98 & 26.02 & 13.27 \\

\singleprompt{Can the sound events described by \{text\} be heard in this audio?}
& 20.00 & 21.50 & 8.00 & 20.92 & 21.43 & 6.12 \\

\singleprompt{Do the sound events described by \{text\} occur in the audio?}
& 21.75 & 22.75 & 9.75 & 22.96 & 19.90 & 7.65 \\

\bottomrule
\end{tabularx}

\caption{
Prompt sensitivity analysis for AQAScore variants on the CompA benchmark.
The placeholder \texttt{\{text\}} is replaced with the corresponding target text prompt.
(Unit: \%).
}
\label{tab:prompt-sensitive-compa}
\end{table*}

%% file: tables/prompt_sensitive_baton.tex
\begin{table*}[ht]
\centering
\small
\begin{tabular}{lcccc}
\toprule
{Prompt Templates}
& \multicolumn{2}{c}{{Baton}} 
& \multicolumn{2}{c}{{Baton-Pair}} \\
\cmidrule(lr){2-3} \cmidrule(lr){4-5}
& AUC (2) & AUC (3)
& Acc. (2) & Acc. (3) \\
\midrule

\multicolumn{5}{l}{{AQAScore (Qwen2.5-Omni-3B)}} \\
\textit{Does the audio contain the sound events described by the text: \{text\}?}
& {0.68} & 0.61 & 70.3 & 69.1 \\
\textit{Is this audio clip containing the sound events described by the text: \{text\}?}
& {0.68} & 0.61 & 71.7 & 68.6 \\
\textit{Does this recording include the sound events described by the text: \{text\}?}
& {0.68} & {0.62} & {73.2} & 69.6 \\
\textit{Does this audio include the sound events described by the text: \{text\}?}
& {0.68} & {0.62} & 71.2 & 68.3 \\
\textit{Does the audio contain the sound events mentioned in the text: \{text\}?}
& {0.68} & {0.62} & 71.7 & {69.8} \\
\textit{Are the sound events described by the text: \{text\} heard in the audio?}
& 0.67 & {0.62} & {73.3} & {70.7} \\
\textit{Are the sound events described in the text: \{text\} present in this audio?}
& {0.68} & 0.61 & 72.2 & 70.4 \\
\textit{Can the sound events described by the text: \{text\} be heard in this audio?}
& 0.67 & 0.61 & 73.0 & {72.0} \\
\textit{Do the sound events described by the text: \{text\} occur in the audio?}
& 0.65 & {0.62} & 72.2 & 71.5 \\

\midrule
\multicolumn{5}{l}{{AQAScore (Qwen2.5-Omni-7B)}} \\
\textit{Does the audio contain the sound events described by the text: \{text\}?}
& {0.69} & {0.65} & {71.0} & 67.8 \\
\textit{Is this audio clip containing the sound events described by the text: \{text\}?}
& 0.68 & {0.65} & 70.2 & 67.2 \\
\textit{Does this recording include the sound events described by the text: \{text\}?}
& {0.69} & {0.65} & {71.0} & {68.0} \\
\textit{Does this audio include the sound events described by the text: \{text\}?}
& {0.69} & {0.65} & 70.8 & 66.9 \\
\textit{Does the audio contain the sound events mentioned in the text: \{text\}?}
& {0.69} & {0.65} & {71.0} & 67.7 \\
\textit{Are the sound events described by the text: \{text\} heard in the audio?}
& 0.67 & {0.65} & 70.3 & 67.7 \\
\textit{Are the sound events described in the text: \{text\} present in this audio?}
& 0.67 & {0.65} & 70.7 & {68.3} \\
\textit{Can the sound events described by the text: \{text\} be heard in this audio?}
& 0.67 & 0.63 & 70.1 & 67.7 \\
\textit{Do the sound events described by the text: \{text\} occur in the audio?}
& 0.67 & 0.64 & 69.5 & 68.2 \\

\midrule
\multicolumn{5}{l}{{AQAScore (AF3)}} \\
\textit{Does the audio contain the sound events described by the text: \{text\}?}
& {0.64} & 0.60 & 66.3 & 63.2 \\
\textit{Is this audio clip containing the sound events described by the text: \{text\}?}
& 0.63 & 0.59 & {67.2} & 62.6 \\
\textit{Does this recording include the sound events described by the text: \{text\}?}
& {0.64} & 0.60 & {67.7} & 62.6 \\
\textit{Does this audio include the sound events described by the text: \{text\}?}
& {0.64} & 0.60 & {67.7} & 62.5 \\
\textit{Does the audio contain the sound events mentioned in the text: \{text\}?}
& {0.64} & {0.61} & {67.8} & {63.2} \\
\textit{Are the sound events described by the text: \{text\} heard in the audio?}
& 0.63 & 0.59 & 66.4 & 60.1 \\
\textit{Are the sound events described in the text: \{text\} present in this audio?}
& 0.63 & 0.58 & 65.5 & 60.9 \\
\textit{Can the sound events described by the text: \{text\} be heard in this audio?}
& 0.61 & 0.57 & 63.1 & 58.5 \\
\textit{Do the sound events described by the text: \{text\} occur in the audio?}
& 0.62 & 0.58 & 65.4 & 61.2 \\

\midrule
\multicolumn{5}{l}{{AQAScore (AF3-Think)}} \\
\textit{Does the audio contain the sound events described by the text: \{text\}?}
& {0.66} & {0.61} & {65.8} & {62.0} \\
\textit{Is this audio clip containing the sound events described by the text: \{text\}?}
& 0.63 & 0.60 & 65.0 & 61.1 \\
\textit{Does this recording include the sound events described by the text: \{text\}?}
& 0.65 & {0.61} & 65.0 & 60.4 \\
\textit{Does this audio include the sound events described by the text: \{text\}?}
& 0.65 & {0.61} & 65.6 & 60.9 \\
\textit{Does the audio contain the sound events mentioned in the text: \{text\}?}
& {0.67} & {0.62} & {66.3} & {61.4} \\
\textit{Are the sound events described by the text: \{text\} heard in the audio?}
& 0.64 & 0.59 & 63.4 & 61.9 \\
\textit{Are the sound events described in the text: \{text\} present in this audio?}
& 0.63 & 0.59 & 63.6 & 60.4 \\
\textit{Can the sound events described by the text: \{text\} be heard in this audio?}
& 0.63 & 0.58 & 64.3 & 59.8 \\
\textit{Do the sound events described by the text: \{text\} occur in the audio?}
& 0.64 & 0.60 & {65.9} & 60.4 \\

\midrule
\multicolumn{5}{l}{{AQAScore (AF3-Chat)}} \\
\textit{Does the audio contain the sound events described by the text: \{text\}?}
& 0.57 & 0.54 & 63.1 & 60.2 \\
\textit{Is this audio clip containing the sound events described by the text: \{text\}?}
& 0.57 & 0.54 & {64.0} & 59.8 \\
\textit{Does this recording include the sound events described by the text: \{text\}?}
& {0.58} & 0.54 & {64.2} & 59.5 \\
\textit{Does this audio include the sound events described by the text: \{text\}?}
& 0.57 & 0.54 & {64.4} & 58.8 \\
\textit{Does the audio contain the sound events mentioned in the text: \{text\}?}
& 0.57 & {0.55} & {64.4} & {59.7} \\
\textit{Are the sound events described by the text: \{text\} heard in the audio?}
& {0.59} & 0.53 & 62.0 & 58.7 \\
\textit{Are the sound events described in the text: \{text\} present in this audio?}
& {0.59} & 0.54 & 61.2 & 57.7 \\
\textit{Can the sound events described by the text: \{text\} be heard in this audio?}
& {0.58} & 0.52 & 60.5 & 57.1 \\
\textit{Do the sound events described by the text: \{text\} occur in the audio?}
& 0.57 & 0.52 & 61.7 & 58.1 \\

\bottomrule
\end{tabular}
\caption{
Prompt sensitivity analysis of AQAScore variants on Baton and Baton-Pair benchmarks.
The placeholder \texttt{\{text\}} is replaced with the corresponding target text prompt.
}
\label{tab:prompt-sensitive-baton}
\end{table*}